# Synthesis and Characterization of a π-Extended Clar's Goblet

Shantanu Mishra,[1,5,‡,*] Manuel Vilas-Varela,[2,‡] Igor Rončević,[3] Fabian Paschke,[1] Florian Albrecht,[1] Leo Gross,[1,*] and Diego Peña,[2,4,*]

[1]IBM Research Europe – Zurich, 8803 Rüschlikon, Switzerland.

[2]Center for Research in Biological Chemistry and Molecular Materials (CiQUS) and Department of Organic Chemistry, University of Santiago de Compostela, 15782 Santiago de Compostela, Spain.

[3]Department of Chemistry, University of Manchester, Oxford Road, Manchester M13 9PL, United Kingdom.

[4]Oportunius, Galician Innovation Agency (GAIN), 15702 Santiago de Compostela, Spain.

[5]Present address: Department of Physics, Chalmers University of Technology, 412 96 Gothenburg, Sweden.

**ABSTRACT:** Concealed non-Kekulé polybenzenoid hydrocarbons have no sublattice imbalance yet cannot be assigned a classical Kekulé structure, leading to an open-shell ground state with potential application in organic spintronics. They constitute an exceedingly small fraction of the total number of polybenzenoid hydrocarbons that can be constructed for a given number of benzenoid rings, and their synthesis remains challenging. The archetype of such a system is Clar's goblet ($C_{38}H_{18}$), a diradical proposed by Erich Clar in 1972 and recently synthesized on a Au(111) surface. Here, we report the synthesis of a π-extended Clar's goblet ($C_{76}H_{26}$), a tetraradical concealed non-Kekulé polybenzenoid hydrocarbon, by a combined in-solution and on-surface synthetic approach. By means of low-temperature scanning tunneling microscopy and atomic force microscopy, we characterized individual molecules adsorbed on a Cu(111) surface. We provide insights into the electronic properties of this elusive molecule, including the many-body nature of its ground and excited states, by mean-field and multiconfigurational quantum chemistry calculations.

Polybenzenoid hydrocarbons (PBHs) may be broadly distinguished on the basis of the presence or absence of Kekulé valence structures, with important implications for their chemical and electronic properties, such as reactivity, aromaticity and, relevant to the context of the present work, magnetism.[1–3] Emergence of magnetism in PBHs may be understood from the interplay of two competing phenomena, namely, (a) the intramolecular hybridization, which drives a closed-shell electronic structure consisting of bonding and antibonding π-orbital pairs in the electronic energy spectrum, such as the highest occupied and lowest unoccupied molecular orbitals (HOMO and LUMO, respectively), and (b) the Coulomb repulsion that penalizes double occupation of an orbital, and drives an open-shell electronic structure with singly occupied molecular orbitals (SOMOs). In this simple picture, Kekulé PBHs exhibit a finite hybridization-induced gap between the frontier molecular orbitals, and therefore a critical value of Coulomb repulsion or a critical system size (that governs the hybridization-induced gap for a homologous series of PBHs) is required for the open-shell solution to become the ground state. In contrast, in non-Kekulé PBHs, it is impossible to pair all $p_z$ electrons into π bonds, and such molecules always contain unpaired electrons. A typical example of non-Kekulé PBHs is the family of [*n*]triangulenes, which are triangular PBHs containing *n* benzenoid rings along each edge. Figure 1a shows the chemical structure of [3]triangulene, which contains two unpaired electrons. In non-Kekulé PBHs such as [*n*]triangulenes, the hybridization-induced gap between the frontier molecular orbitals is negligible, and inclusion of an arbitrarily small Coulomb repulsion triggers spin polarization resulting in an open-shell ground state. The underlying reason for the non-Kekulé structure of [*n*]triangulenes is an inherent sublattice imbalance in the bipartite honeycomb lattice.

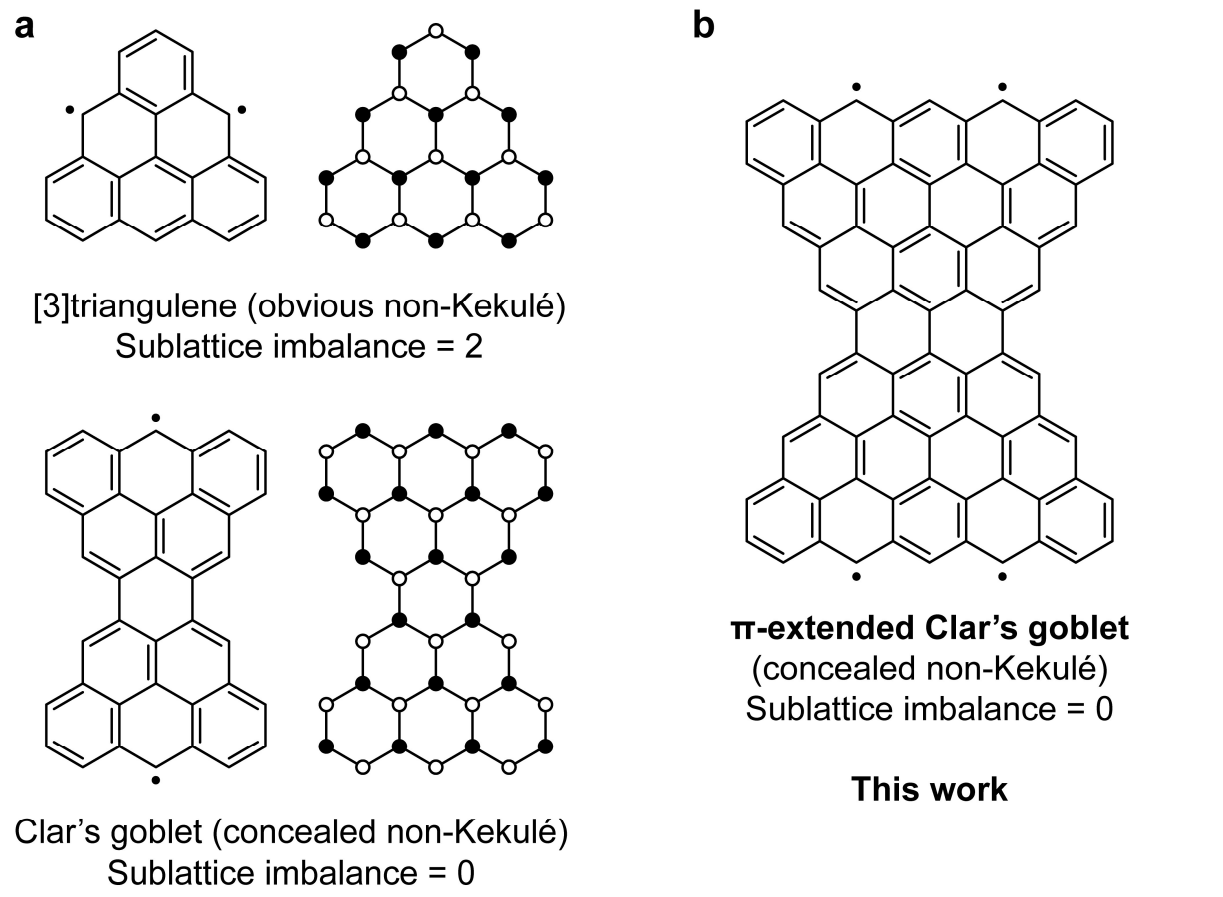

**Figure 1.** (a) Chemical structures and sublattice representations of [3]triangulene and Clar's goblet. Filled and empty circles denote the two sublattices. (b) Chemical structure of the π-extended Clar's goblet.

**Scheme 1. Synthetic route toward ECG**[a]

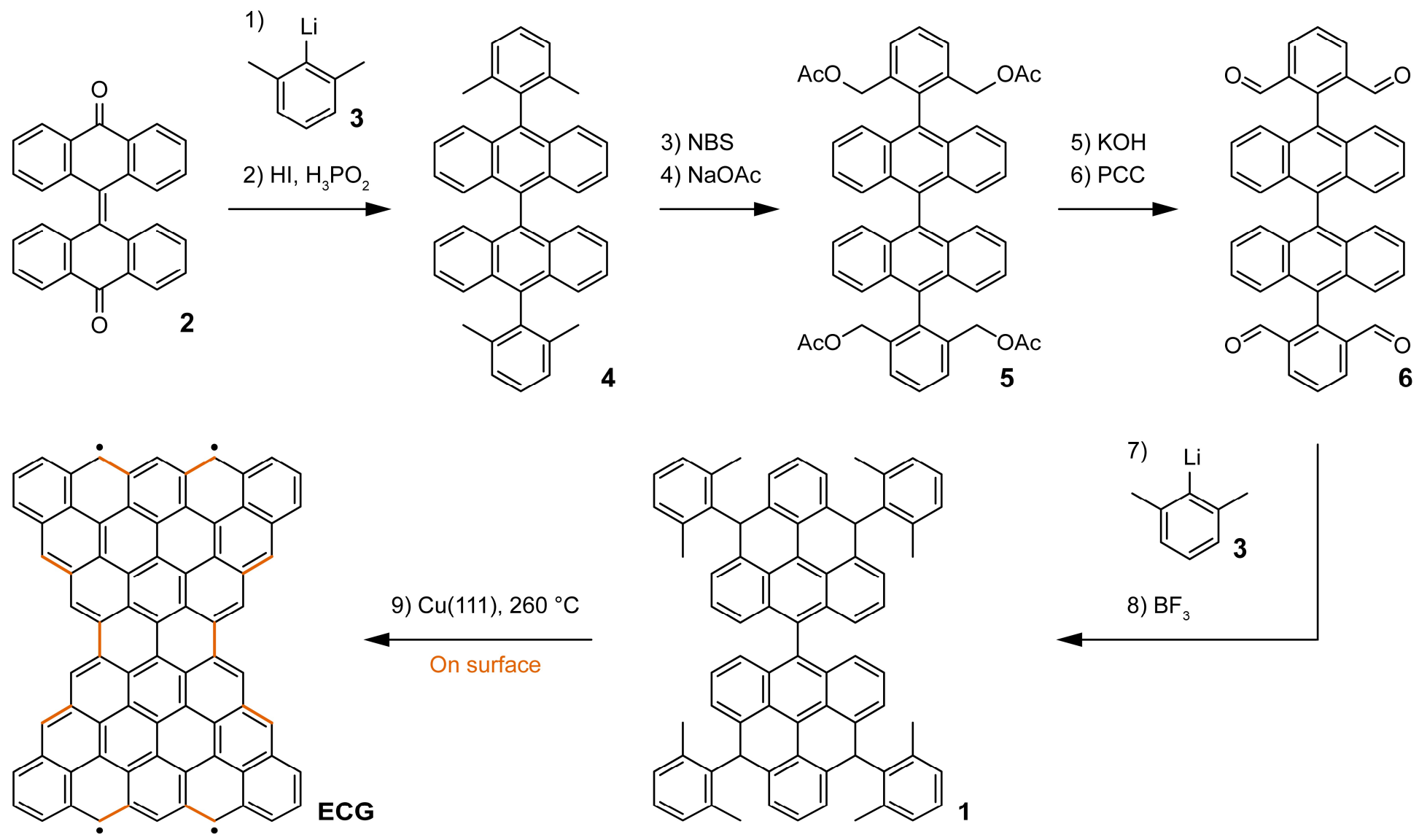

[a]Highlighted bonds in **ECG** are formed from on-surface reactions of **1**.

[*n*]triangulenes have a sublattice imbalance of *n*–1, and therefore a ground state total spin quantum number of (*n*–1)/2 from Ovchinnikov's rule.[4,5] In literature, such molecules have been referred to as *obvious* non-Kekulé PBHs,[6] and many of them have recently been synthesized both in solution and on surfaces, for example, [*n*]triangulenes with *n* = 2–7,[7–13] aza-[*n*]triangulenes with *n* = 3 and 5,[14,15] a [7]triangulene coronoid,[16] olympicene,[17,18] and other [*n*]triangulene-based molecules.[19] However, sublattice imbalance is not a necessary condition to generate non-Kekulé PBHs, and there are examples of so-called *concealed* non-Kekulé PBHs[6] that cannot be assigned a classical Kekulé structure despite the absence of sublattice imbalance. Ovchinnikov's rule predicts a singlet ground state for concealed non-Kekulé PBHs, owing to a lack of sublattice imbalance. The first such system was proposed by Clar in 1972, the eponymous Clar's goblet ($C_{38}H_{18}$, Fig. 1a).[20,21] In 1974, Gutman showed that concealed non-Kekulé PBHs can only be constructed for $h \geq 11$, where *h* denotes the number of benzenoid rings in a PBH.[22] It was further shown that concealed non-Kekulé PBHs constitute a small fraction of the total number of PBHs that can be constructed for a given *h*,[23] with abundancies < 0.1% for $h \leq 14$. To the best of our knowledge, only three concealed non-Kekulé PBHs have been synthesized to date, namely, unsubstituted Clar's goblet reported by Mishra et al.[24] and its substituted derivative reported by Jiao et al.,[25] a butterfly-shaped PBH reported by Song et al.[26], and a phenalene-perylene fused system reported by Imran et al.[27] The rarity of concealed non-Kekulé PBHs, along with the proposed application of these systems as components of spintronic devices,[28–30] make them interesting synthetic targets in organic chemistry. Here, we report a combined in-solution and on-surface synthesis of a π-extended Clar's goblet ($C_{76}H_{26}$, **ECG**; Fig. 1b) on Cu(111), and its characterization by means of scanning tunneling microscopy (STM) and atomic force microscopy (AFM) operating under ultra-high vacuum and at a temperature of 5 K. Through mean-field and multiconfigurational quantum chemistry calculations, we provide insights into the electronic structure of **ECG** in the gas phase, including its tetraradical character, and its many-body ground and excited spin states.

We envisioned the synthesis of **ECG** by surface-catalyzed cyclodehydrogenation and oxidation reactions of the precursor 8,8′,12,12′-tetrakis(2,6-dimethylphenyl)-8,8′,12,12′-tetrahydro-4,4′-bidibenzo[*cd*,*mn*]pyrene (**1**, Scheme 1). The on-surface reactions would involve the formation of 10 new carbon-carbon bonds and the cleavage of 32 carbon-hydrogen bonds. Compound **1** was obtained by solution chemistry in eight steps starting from bianthrone **2**. First, the addition of two equivalents of organolithium derivative **3**, followed by reduction, led to the formation of substituted bianthracene **4** in 76% yield. Then, four-fold bromination with *N*-bromosuccinimide (NBS), followed by treatment with sodium acetate (NaOAc), afforded **5** in 33% yield. Ester hydrolysis of **5** under basic conditions, followed by oxidation with pyridinium chlorochromate (PCC), led to the formation of tetra-aldehyde **6** in 49% yield. Finally, addition of four equivalents of **3**, followed by $BF_3$-promoted four-fold intramolecular Friedel-Crafts reaction led to the isolation of **1** as a mixture of diastereoisomers in 70% yield. Details on solution synthesis and characterization are reported in Figs. S1–S11.

We first attempted the on-surface synthesis of **ECG** on Au(111), which is the least reactive of all coinage metal surfaces, and where many open-shell PBHs are shown to be weakly physisorbed without considerable alteration of their gas-phase electronic structure. A submonolayer coverage of **1** was deposited on a Au(111) surface held at 10 K, and the surface was annealed to 600 K for 5 minutes to promote the on-surface reactions. Following this protocol, we observed a universal loss of methyl groups from **1**, which led to the formation of pentagonal rings upon cyclodehydrogenation reactions (Fig. S12). Other undesired reactions on Au(111) included the loss of one or more xylyl groups, fragmentation of the precursor into 4,8-bis(2,6-dimethylphenyl)-4,8-dihydrobenzo[*cd*,*mn*]pyrene units, and intermolecular dehydrogenative cross-coupling reactions. We did not find **ECG** from imaging more than 100 isolated molecules on the surface. We then attempted the synthesis of **ECG** on Cu(111), with the rationale that the higher catalytic activity of Cu(111) compared to Au(111) would require lower temperatures to trigger cyclization reactions, potentially avoiding the problem of methyl cleavage. Moreover, the higher diffusion barrier of molecules on Cu compared to Au could also reduce intermolecular reactions, as was observed in previous studies.[10,11,31] Figure 2a presents an overview STM image after annealing a submonolayer coverage of **1** on Cu(111) at 530 K for 5 minutes, wherein the surface was covered by mostly isolated molecules. Approximately 25% of the molecules on the surface corresponded to **ECG**. Figure 2b, c present AFM images of **ECG** on Cu(111), showing the expected atomic structure of the molecule. In Fig. S13, we present AFM images of molecules on the surface that do not correspond to **ECG**, resulting from loss or migration of methyl or xylyl groups, incorporation of carbon atoms, or precursor fragmentation. In AFM imaging (Fig. 2b and Fig. S14), the central pyrene moiety of **ECG** is imaged brighter (that is, with a more positive frequency shift $\Delta f$) because of stronger repulsive forces, whereas the benzenoid rings along the long zigzag edges appear darker. This may indicate a non-planar adsorption conformation of **ECG**, wherein the carbon atoms at the zigzag edges (that harbor the highest spin density as shown in Fig. 3) strongly interact with the underlying Cu atoms and are pulled toward the surface. Density functional theory calculations of **ECG** on Cu(111) confirm this scenario (Fig. S15). Our observations are in line with a previous study of [7]triangulene on Cu(111) where the molecule was found to chemisorb on the Cu surface with a non-planar adsorption geometry.[11]

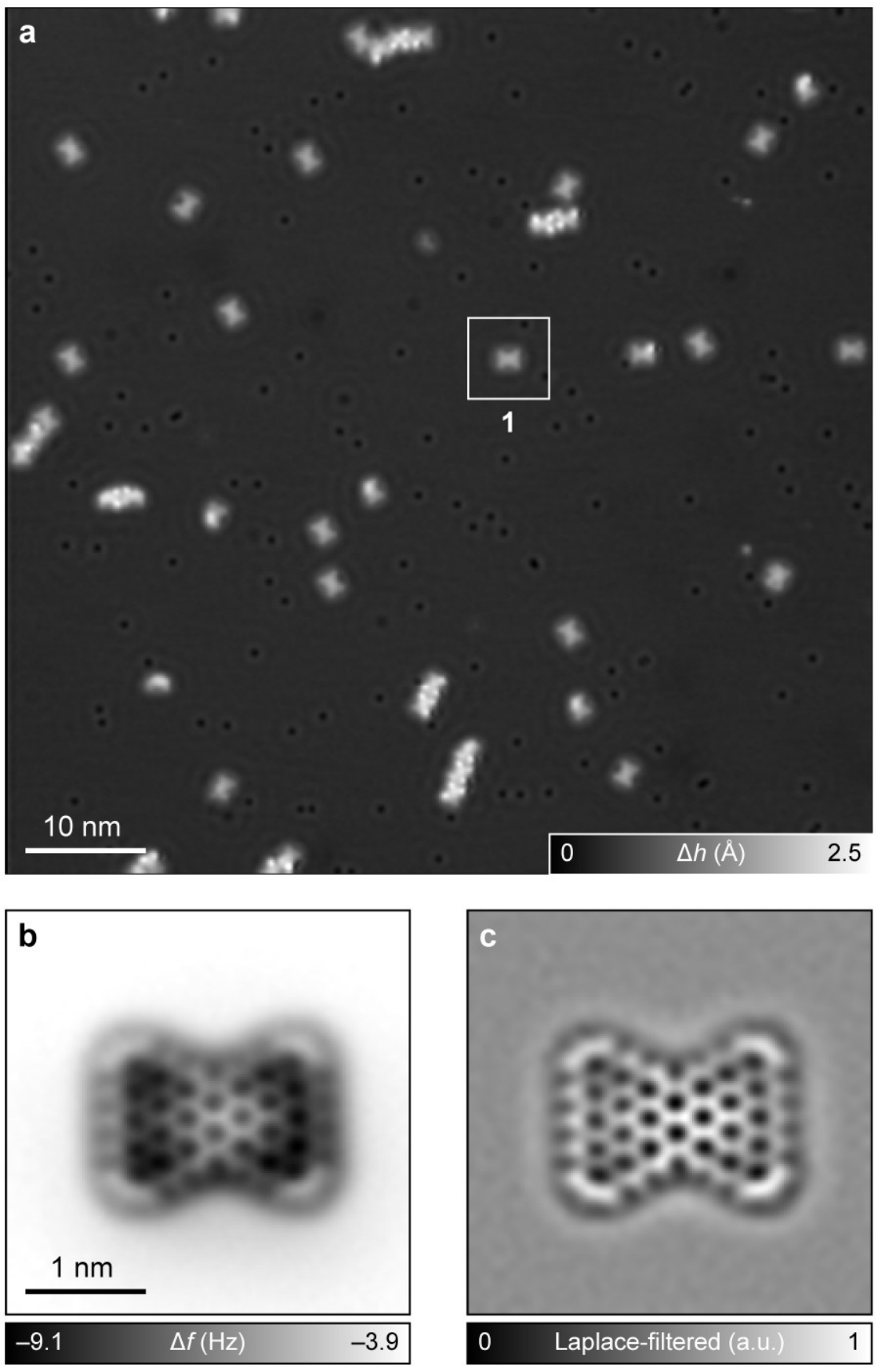

**Figure 2.** On-surface synthesis and structural characterization of **ECG**. (a) Overview STM image after annealing **1** on Cu(111). Scanning parameters: bias voltage $V$ = 0.2 V, tunneling current $I$ = 0.5 pA. $\Delta h$ denotes the tip height. (b) AFM image of the highlighted **ECG** molecule in (a). The tip was approached by 2.8 Å from the STM setpoint $V$ = 0.2 V, $I$ = 0.5 pA on Cu(111). (c) Corresponding Laplace-filtered AFM image. a.u. denotes arbitrary units.

We now discuss the electronic properties of **ECG** by both mean-field and multiconfigurational calculations. We start by performing nearest-neighbor tight-binding calculations, considering only $p_z$ orbitals at the carbon sites, which provides an intuitive (albeit overly simplistic) picture of the electronic structure of **ECG**. Figure 3a shows the tight-binding energy spectrum of **ECG**, where the important feature is the presence of four states at zero energy (zero modes) that are populated by four electrons. Away from the zero modes, one finds a series of bonding and anti-bonding orbital pairs (the first of such pairs is indicated as HOMO and LUMO). We then included electronic correlations in **ECG** via the mean-field Hubbard (MFH) approximation, where an intra-atomic Coulomb repulsion term is added to the tight-binding Hamiltonian to account for the energy cost of having a molecular orbital doubly occupied. The MFH solution predicted an open-shell singlet ground state of **ECG**, in agreement with Ovchinnikov's rule and previous studies.[28,30] Figure 3a, b shows the MFH spectrum and spin polarization plot of **ECG** in the open-shell singlet state. The degeneracy of the zero modes is now lifted by spin polarization, resulting in the formation of four SOMOs (and the corresponding unoccupied molecular orbitals, SUMOs) labeled $\psi_1$–$\psi_4$. The SOMOs are sublattice polarized, with SOMOs of opposite spins localized on different sublattices and on opposite halves of **ECG**.[24,28] To obtain more accurate insights into the electronic structure of **ECG**, we performed calculations accounting for the many-body nature of the ground and excited states (see Supporting Information for details).[32] Briefly, we performed complete active space self-consistent field (CASSCF) calculations wherein the occupation of the orbitals corresponding to the four zero modes is allowed to vary, while the occupations of the orbitals lower or higher in energy are frozen. The active space thus consists of four electrons in the four zero modes, that is, CASSCF(4,4). The CASSCF(4,4) calculations confirmed the open-shell singlet ground state of **ECG**. To accurately determine excited-state energies and exchange interactions, we employed the difference-dedicated configuration interaction (DDCI) method.[33] DDCI improves upon CASSCF by building a CI space that includes all single and double excitations involving the active space. Figure 3c, d shows the CASSCF(4,4)-DDCI-calculated spectrum of the ground and excited states of **ECG**. Relative to the (open-shell) singlet ground state, the first excited state is a triplet that is 8 meV higher in energy, followed by the second excited state at 11 meV, which is a quintet. Further up in energy are two nearly degenerate triplet states at 381 meV, and a singlet state at 725 meV. Most of these states exhibit a strong many-body character that cannot be captured by mean-field theories (see Fig. S16 and accompanying note). We also rationalized the many-body states shown in Fig. 3c, d in terms of model spin Hamiltonians with linear and non-linear exchanges, and derived the relevant exchange interactions (Fig. S17). Finally, we analyzed the aromaticity of **ECG** by calculating induced currents and nucleus-independent chemical shifts (NICS) in the open-shell singlet state (Fig. 3e). The induced current map reveals the presence of two diatropic ring currents in the two halves of **ECG**, which meet in the centre to produce a strong paratropic current. NICS values confirm aromatic spin-bearing dibenzo[*bc*,*pq*]ovalenyl moieties and an anti-aromatic central napthalene moiety, similar to Clar's goblet[34] and in agreement with the Clar sextet distribution of **ECG** (Fig. 1b).

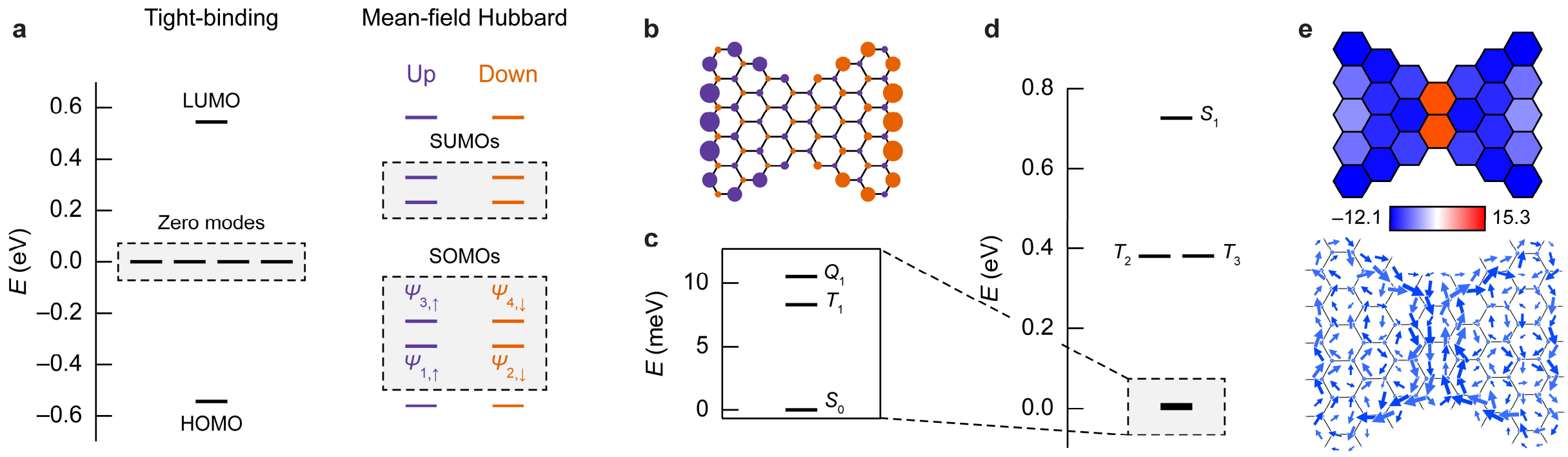

**Figure 3**. Theoretical electronic characterization of **ECG**. (a) Nearest-neighbor tight-binding (left) and MFH (right) spectra of **ECG**. (b) MFH spin polarization plot of **ECG**, expressed as the difference in the mean populations of spin up and down electrons. Size and color of the circles denote the value (the largest and smallest absolute values are 0.295 and 0.017 electrons) and sign of spin polarization, respectively. (c, d) CASSCF(4,4)-DDCI spectrum of **ECG**. The low-energy manifold of spin states is shown in (c). The labels $S$, $T$ and $Q$ denote singlet, triplet and quintet states, respectively, and 0, 1 and 2 denote the energetic order (low to high, respectively) of states of a given multiplicity. (e) NICS(1)$_{zz}$ (top) and induced current (bottom) maps of **ECG**. Negative (positive) NICS(1)$_{zz}$ values indicate local aromaticity (antiaromaticity).

In summary, by in-solution and on-surface chemistry, we synthesized a π-extended Clar's goblet (**ECG**) on Cu(111), and characterized its chemical structure by STM and AFM imaging. We analyzed the electronic structure of **ECG** through mean-field and multiconfigurational quantum chemistry calculations. We found that **ECG** is a tetraradical system with a singlet ground state, and close-lying triplet and quintet excited states, which exhibit strong many-body character. Our study demonstrates synthetic access to a rare class of polyradical conjugated systems with proposed application in spintronics.

## ASSOCIATED CONTENT

**Supporting Information**. Methods, additional STM and AFM data, and additional calculations.

## AUTHOR INFORMATION

### Corresponding Author

*shantanu.mishra@chalmers.se
*diego.pena@usc.es
*LGR@zurich.ibm.com

### Author Contributions

‡These authors contributed equally.

S.M. and L.G. performed on-surface synthesis and scanning probe microscopy measurements. M.V.-V. and D.P. synthesized and characterized the precursor molecule in solution. S.M. performed the tight-binding calculations. I.R. performed density functional theory, nucleus-independent chemical shift, and multiconfigurational quantum chemistry calculations. S.M. wrote the manuscript, with contributions from all authors.

### Notes

The authors declare no competing financial interest.

## ACKNOWLEDGMENT

This study has received funding from the European Union project SPRING (grant number 863098), the European Research Council Synergy grant MolDAM (grant number 951519), the Spanish Agencia Estatal de Investigación (grant number PID2022-140845OB-C62), Xunta de Galicia (Centro de Investigación de Galicia accreditation 2019–2022, grant number ED431G 2019/03), and the European Regional Development Fund. I.R. acknowledges support from The University of Manchester. Computational support was provided by Research IT and the Computational Shared Facility at The University of Manchester.

# Supporting Information

## Synthesis and Characterization of a π-Extended Clar's Goblet

Shantanu Mishra,[1,5] Manuel Vilas-Varela,[2] Igor Rončević,[3] Fabian Paschke,[1] Florian Albrecht,[1] Leo Gross,[1] and Diego Peña,[2,4]

[1]IBM Research Europe – Zurich, 8803 Rüschlikon, Switzerland.

[2]Center for Research in Biological Chemistry and Molecular Materials (CiQUS) and Department of Organic Chemistry, University of Santiago de Compostela, 15782 Santiago de Compostela, Spain.

[3]Department of Chemistry, University of Manchester, Oxford Road, Manchester M13 9PL, United Kingdom.

[4]Oportunius, Galician Innovation Agency (GAIN), 15702 Santiago de Compostela, Spain.

[5]Present address: Department of Physics, Chalmers University of Technology, 412 96 Gothenburg, Sweden.

**Contents**

# 1. Methods

## 1.1. Solution synthesis and characterization.

Starting materials were purchased reagent grade from TCI and Sigma-Aldrich and used without further purification. Bianthrone **2** was obtained following reported procedures.[1] Organolithium reagent **3** was prepared by treatment of 2-iodo-1,3-dimethylbenzene with *n*-butyllithium. All reactions were carried out in flame-dried glassware under an inert atmosphere of purified Ar using Schlenk techniques. Thin-layer chromatography was performed on Silica Gel 60 F-254 plates (Merck). Column chromatography was performed on silica gel (40-60 µm). Nuclear magnetic resonance (NMR) spectra were recorded on a Bruker Varian Inova 500 spectrometer. Mass spectra (MS) were recorded on a Bruker MicroTOF spectrometer.

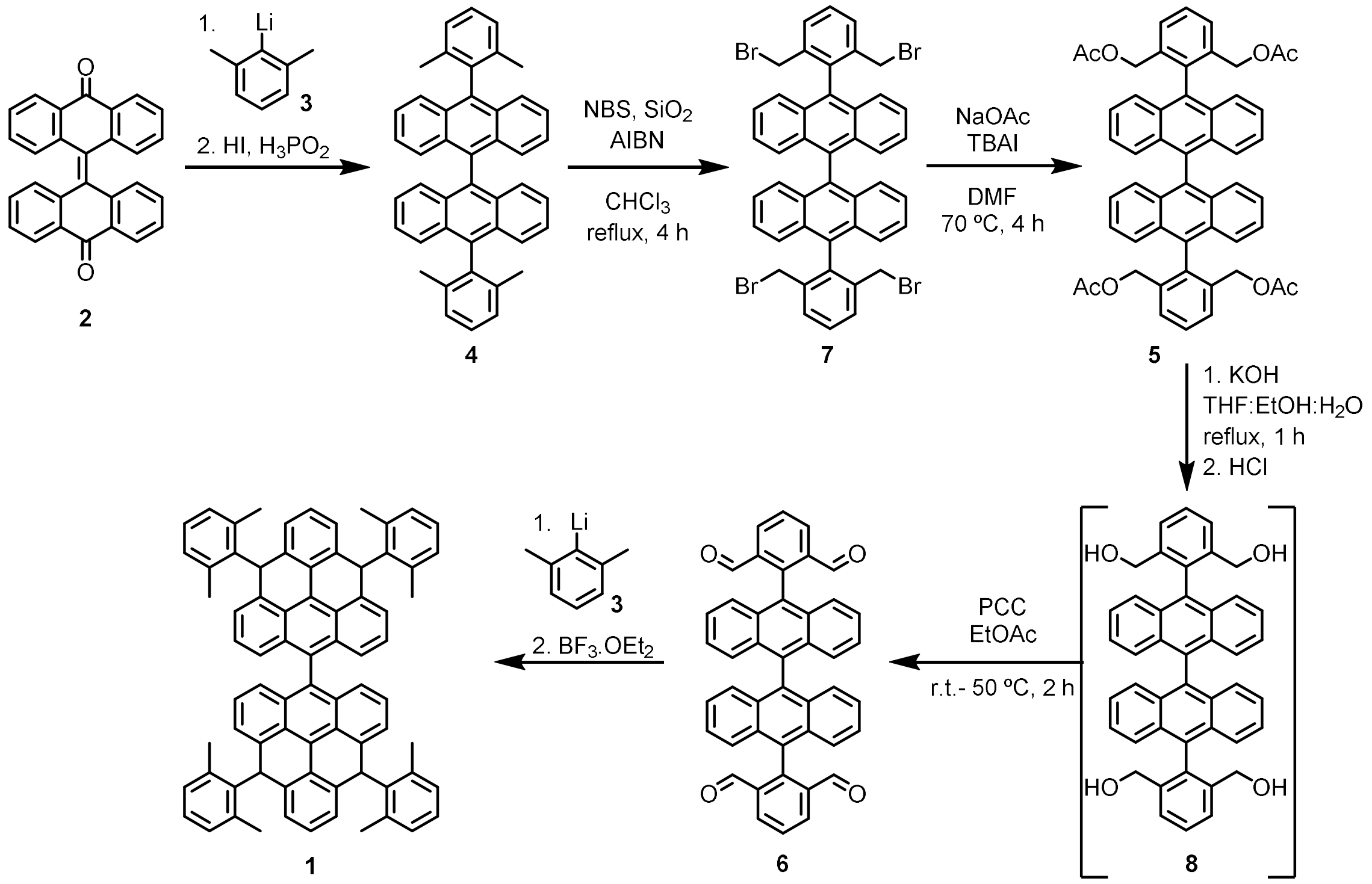

**Fig. S1.** Synthetic route toward **1**.

### Synthesis of 10,10'-bis(2,6-dimethylphenyl)-9,9'-bianthracene (4)

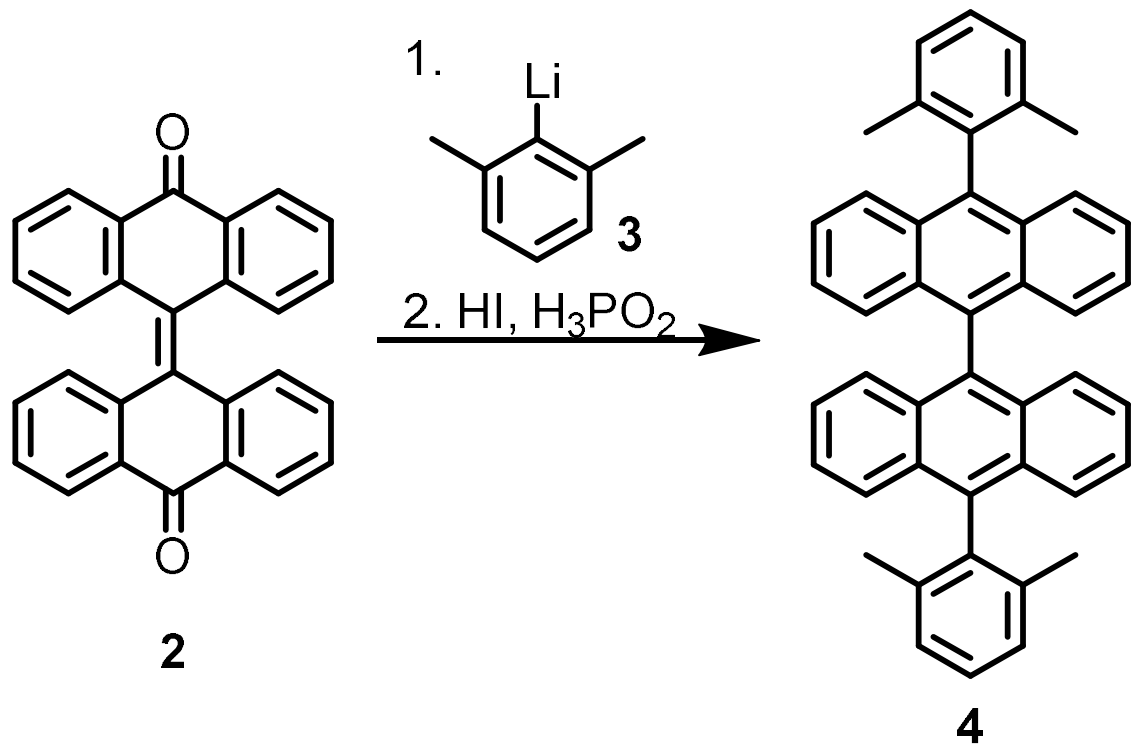

**Fig. S2.** Synthesis of **4**.

Over a solution of **3** (13.0 mmol) in $Et_2O$ (60 mL), compound **2** (1.00 g, 2.60 mmol) was added at 0 °C. The resulting mixture was allowed to reach room temperature and stirred for 16 h. Then, AcOH (1.00 mL) was added, and the solvent was removed under reduced pressure. The residue was suspended in AcOH (40 mL). Then, $H_3PO_2$ (20 mL) and HI (4 mL) were added. The resulting suspension was heated at 90 °C for 3 h. After cooling to room temperature, $H_2O$ (100 mL) was added, and the precipitate was filtered and washed with $H_2O$ (2 × 15 mL) and MeOH (2 × 10 mL). The obtained solid was purified by column chromatography ($SiO_2$; hexane:$CH_2Cl_2$ 4:1) to afford **4** (1.11 g, 76%) as a yellow solid. **$^1$H-NMR** (500 MHz, $CDCl_3$) δ: 7.64 (d, *J* = 8.7 Hz, 4H), 7.49 – 7.41 (m, 2H), 7.41 – 7.29 (m, 8H), 7.27 – 7.14 (m, 8H), 2.00 (s, 12H) ppm. **$^{13}$C- NMR** (125 MHz, $CDCl_3$) δ: 138.0 (C), 136.5 (C), 133.2 (C), 131.8 (C), 129.6 (C), 127.9 (CH), 127.7 (CH), 127.4 (CH), 126.4 (CH), 125.8 (CH), 125.7 (CH), 20.5 ($CH_3$) ppm. **MS (APCI)** *m/z* (%): 562 (M+, 100). **HRMS (APCI)**: $C_{44}H_{35}$; calculated: 563.2733, found: 562.2734.

**Synthesis of 10,10'-bis(2,6-bis(bromomethyl)phenyl)-9,9'-bianthracene (7)**

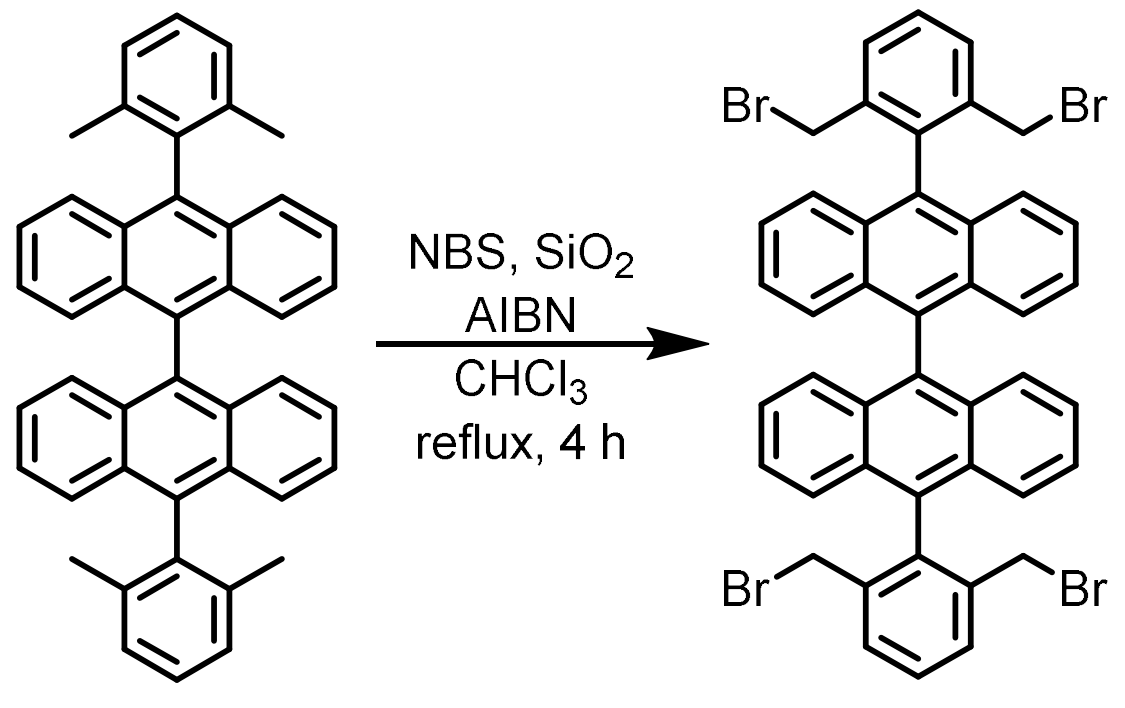

**Fig. S3.** Synthesis of **7**.

Over a mixture of **4** (100 mg, 0.18 mmol), NBS (160 mg, 0.90 mmol) and $SiO_2$ (150 mg) in $CHCl_3$ (10 mL), a catalytic amount of AIBN was added. The mixture was heated at reflux for 4 h. The solvent was removed under reduced pressure and the residue was purified by column chromatography ($SiO_2$; hexane:$CH_2Cl_2$ 4:1 to 2:1) to afford **7** (50 mg, 35%) as a yellowish solid. **$^1$H NMR** (500 MHz, $CDCl_3$) δ: 7.81 (d, *J* = 7.7 Hz, 4H), 7.68 (dd, *J* = 8.4, 6.9 Hz, 2H), 7.56 (d, *J* = 8.7 Hz, 4H), 7.37 (ddd, *J* = 8.7, 5.3, 2.4 Hz, 4H), 7.24 (dd, *J* = 5.9, 4.0 Hz, 8H), 4.19 (s, 8H) ppm. **$^{13}$C NMR** (125 MHz, $CDCl_3$) δ: 138.20 (C), 131.50 (C), 131.30 (CH), 130.30 (C), 129.6 (CH), 127.30 (CH), 126.70 (CH), 126.40 (CH), 126.20 (CH), 31.60 ($CH_2$). ppm. **MS (APCI)** *m/z* (%): 878 (M+, 100), 799 (18). **HRMS (APCI)**: $C_{44}H_{30}Br_4$; calculated: 873.9076, found: 873.9070.

**Synthesis of ([9,9'-bianthracene]-10,10'-diylbis(benzene-2,1,3-triyl))tetrakis(methylene) tetraacetate (5)**

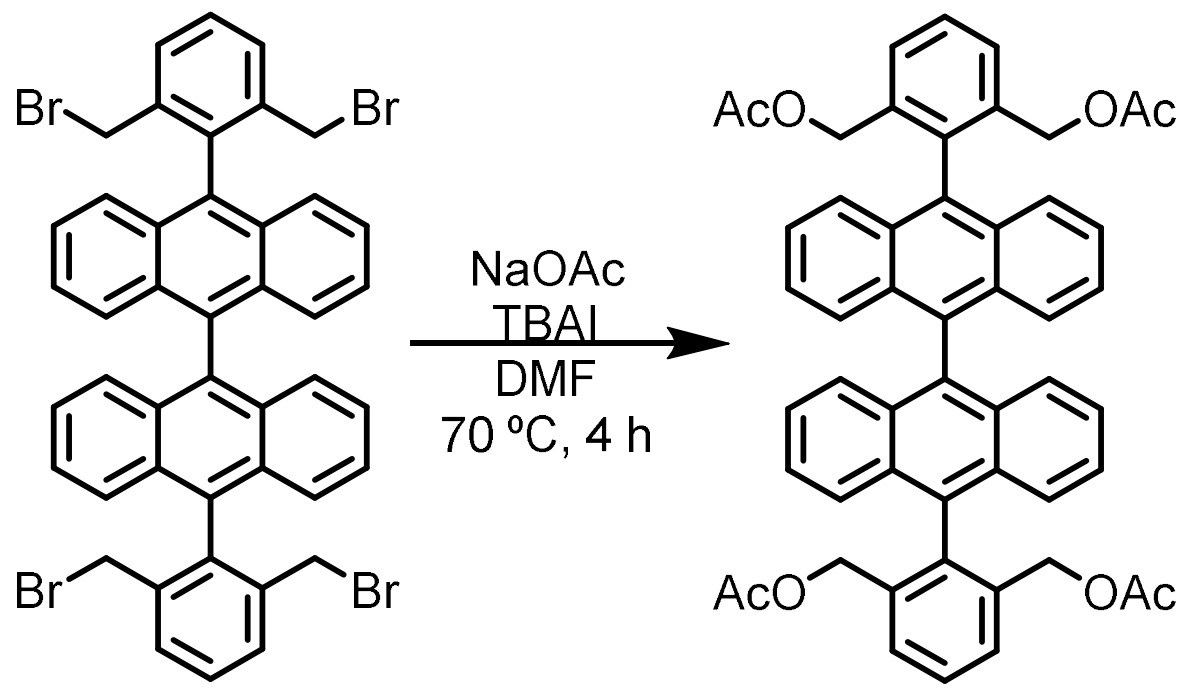

**7** **5**

**Fig. S4.** Synthesis of **5**.

A mixture of **7** (50 mg, 0.06 mmol), NaOAc (50 mg, 0.60 mmol) and TBAI (5 mg, 0.01 mmol) in DMF (5 mL) was heated at 80 °C for 4 h. After cooling to room temperature, $H_2O$ (50 mL) was added, and the mixture was extracted with EtOAc (2 × 15 mL). The combined organic extracts were dried over $MgSO_4$, filtered, and evaporated under reduced pressure. The residue was purified by column chromatography ($SiO_2$; hexane:EtOAc 4:1) to afford **5** (45 mg, 94%) as a yellowish solid. $^1$**H NMR** (500 MHz, $CDCl_3$) δ: 7.70 (q, *J* = 4.9 Hz, 6H), 7.53 (d, *J* = 8.8 Hz, 4H), 7.37 – 7.29 (m, 4H), 7.20 (d, *J* = 3.5 Hz, 8H), 4.78 (s, 8H), 1.50 (S, 12H) ppm. $^{13}$**C NMR** (125 MHz, $CDCl_3$) δ: 170.2 (CO), 138.8 (C), 136.4 (C), 134.5 (C), 131.9 (C), 131.4 (C), 130.1 (C), 129.8 (CH), 128.7 (CH), 127.4 (CH), 126.3 (CH), 126.0 (CH), 64.7 ($CH_2$), 20.1($CH_3$) ppm. **MS (APCI)** *m/z* (%): 794 (M+, 100), 735 (3). **HRMS (APCI)**: $C_{52}H_{42}O_8$; calculated: 794.2874, found: 794.2873.

**Synthesis of 2,2'-([9,9'-bianthracene]-10,10'-diyl)diisophthalaldehyde (6)**

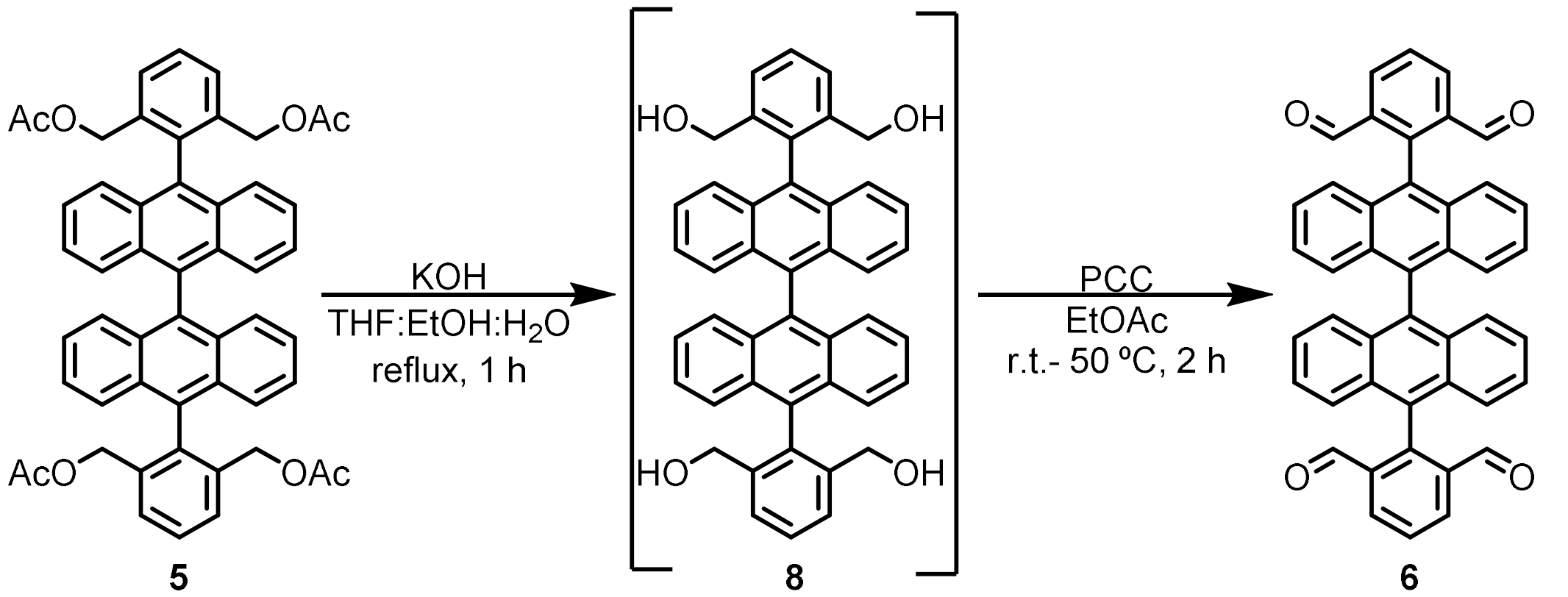

**5** **8** **6**

**Fig. S5.** Synthesis of **6**.

A mixture of **5** (510 mg, 0.64 mmol) and KOH (360 mg, 6.40 mmol) in THF:EtOH:$H_2O$ (2:2:2, 20 mL) was refluxed for 1 h. After cooling to room temperature, $H_2O$ (50 mL) was added, and the pH was adjusted to 5 by addition of $HCl_{(aq)}$. The resulting mixture was extracted with EtOAc (5 × 25 mL). The combined organic extracts were dried over $MgSO_4$, filtered and evaporated under reduced pressure. The crude product **8** was suspended in EtOAc:$CH_2Cl_2$ (3:1, 100 mL), and PCC (690 mg, 3.2 mmol) was portion-wise added at room temperature. After the addition, the mixture was heated at 50 °C for 2 h, cooled to room temperature and quenched by addition of *i*PrOH (5 mL). Solvents were removed under reduced pressure and the residue was purified by column chromatography ($SiO_2$; $CH_2Cl_2$) to afford **6** (192 mg, 49%) as a yellowish solid. $^1$**H NMR** (500 MHz, $CDCl_3$) δ: 9.56 (s, 4H), 8.58 (d, *J* = 7.7 Hz, 4H), 7.97 (t, *J* = 7.7 Hz, 2H), 7.53 – 7.40 (m, 8H), 7.33 – 7.20 (m, 8H) ppm. $^{13}$**C NMR** (125 MHz, $CDCl_3$) δ:

190.7 (CO), 145.8 (C), 136.5 (C), 135.1 (C), 133.3 (CH), 131.9 (C), 130.9 (C), 129.7 (CH), 127.6 (CH), 127.4 (CH), 127.3 (C), 126.8 (CH), 126.2 (CH) ppm. **MS (APCI)** *m/z* (%): 618 (M+, 100). **HRMS (APCI)**: $C_{44}H_{27}O_4$; calculated: 619.1904, found: 619.1881.

**Synthesis of 1**

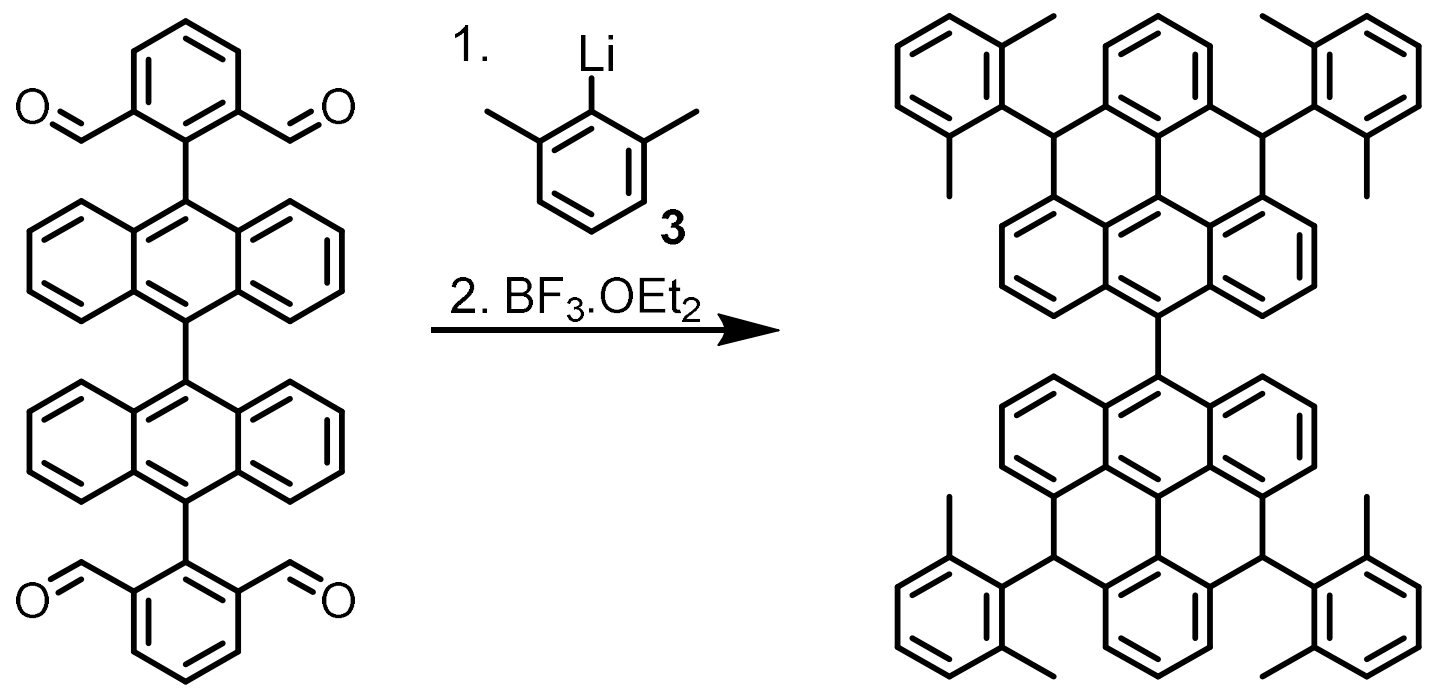

**6** **1**

**Fig. S6.** Synthesis of **1**.

Over a solution of **3** (0.20 mmol) in $Et_2O$ (2 mL), a suspension of **6** (12 mg, 0.02 mmol) in $Et_2O$:THF (1:2, 6 mL) was added, and the resulting mixture was stirred at room temperature for 4 h. Then, $H_2O$ (5 mL) was added, and the mixture was extracted with EtOAc (2 × 10 mL). The combined organic extracts were dried over $MgSO_4$, filtered and evaporated under reduced pressure. The obtained product was dissolved in $CH_2Cl_2$ (3 mL) and $BF_3.OEt_2$ (40 µL) was added at 0 ºC. The resulting mixture was allowed to reach room temperature and stirred for 30 min. Then, $H_2O$ (5 mL) was added, the phases were separated, and the aqueous phase was extracted with $CH_2Cl_2$ (2 × 5 mL). The combined organic extracts were dried over $MgSO_4$, filtered and evaporated under reduced pressure. The residue was purified by column chromatography ($SiO_2$; hexane: $CH_2Cl_2$ 4:1 to 1:1) to afford **1** (5 mg, 70%) as an orange solid (mixture of diastereomers). **$^1$H NMR** (500 MHz, $CDCl_3$) δ: 7.31 – 7.13 (m, 12H), 7.07 – 6.97 (m, 10H), 6.87 (d, *J* = 6.8 Hz, 4H), 6.80 (d, *J* = 7.6 Hz, 4H), 6.72 – 6.46 (m, 4H), 2.68 (m, 12H), 2.03 – 1.65 (m, 12H) ppm. **MS (APCI)** *m/z* (%): 970 (M+, 100), 865 (8). **HRMS (APCI)**: $C_{76}H_{58}$; calculated: 970.4533, found: 970.4547.

**[1]H and [13]C NMR spectra**

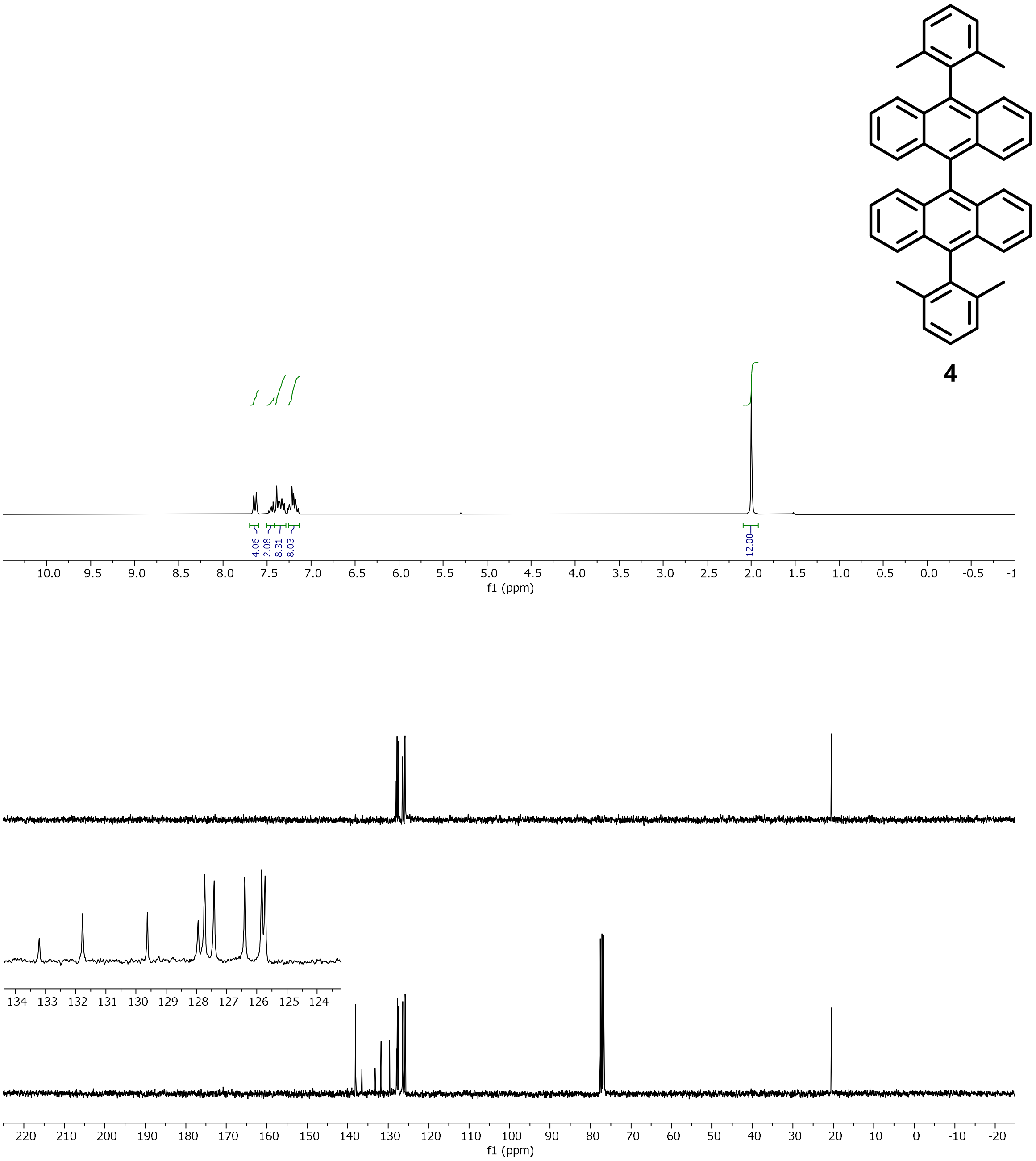

**Fig. S7.** $^{1}H$ and $^{13}C$ NMR spectra of **4**.

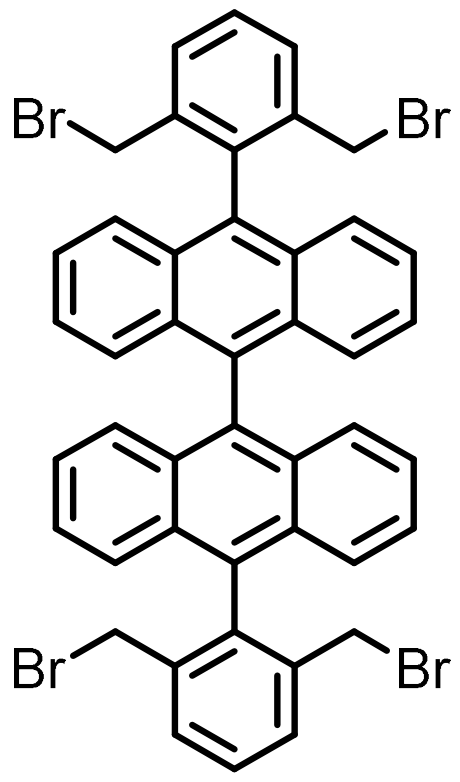

**7**

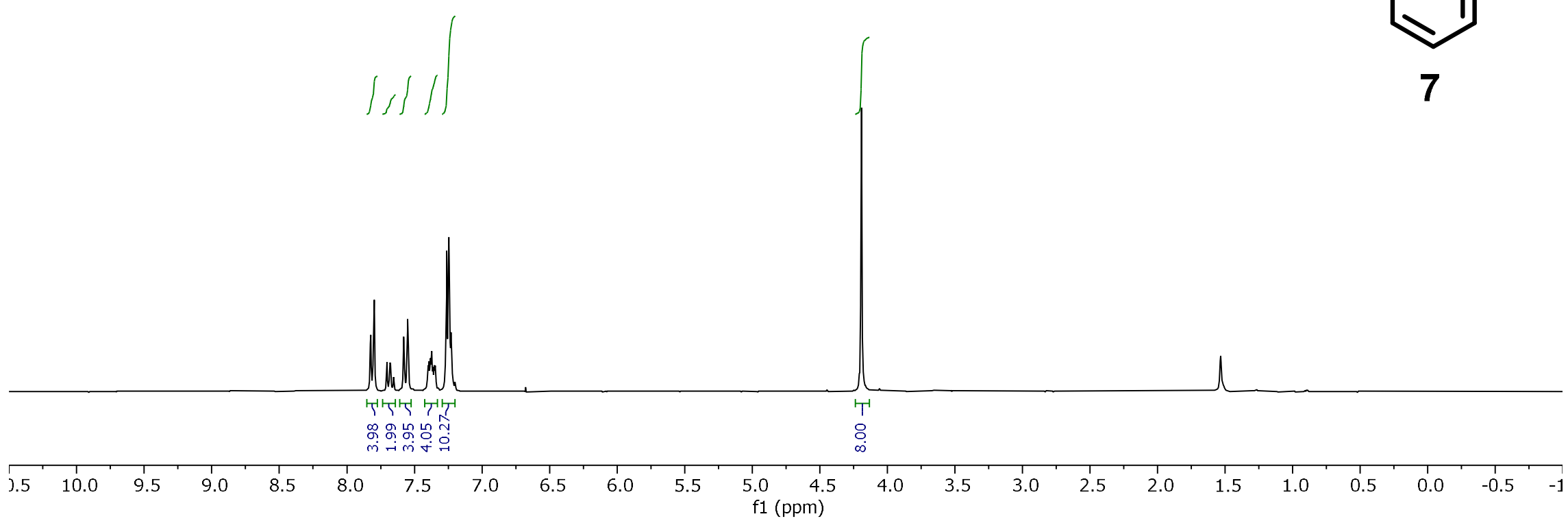

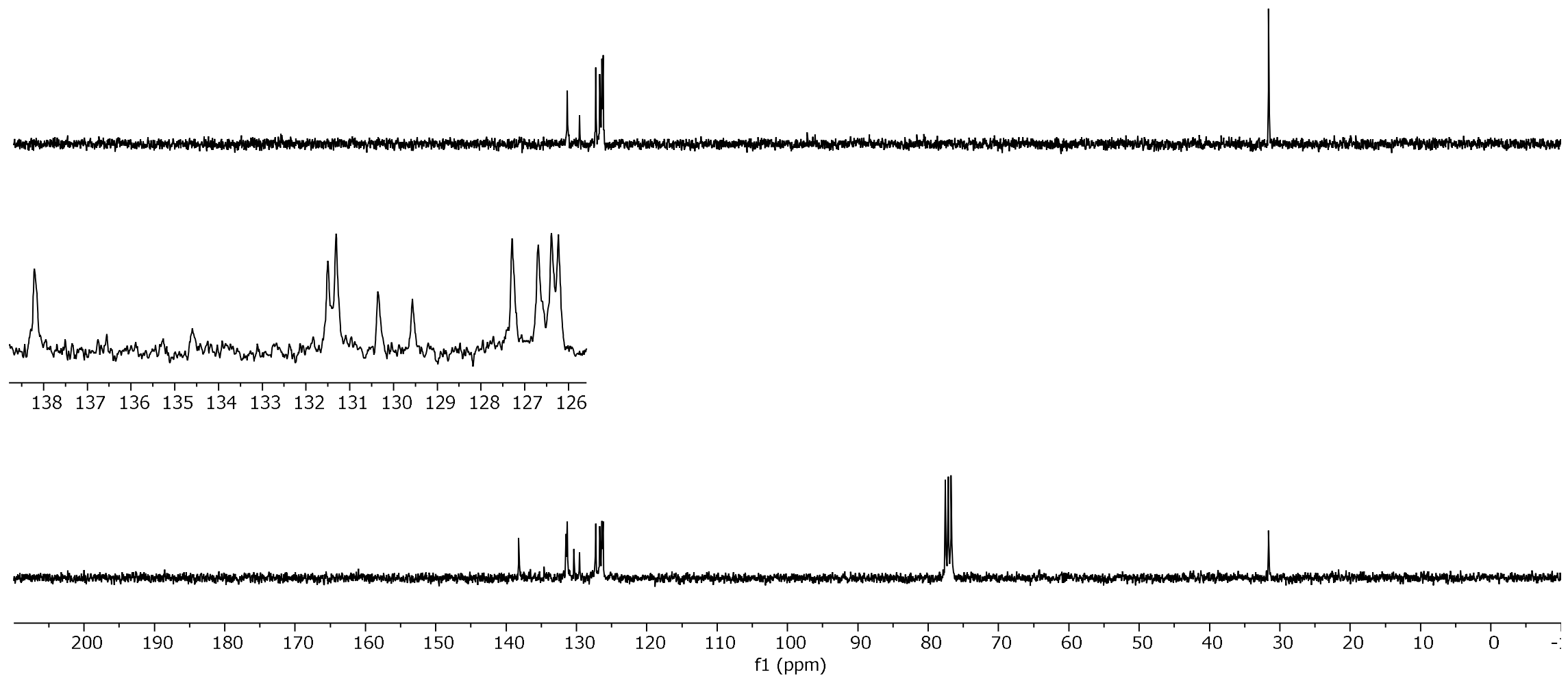

**Fig. S8.** $^{1}H$ and $^{13}C$ NMR spectra of **7**.

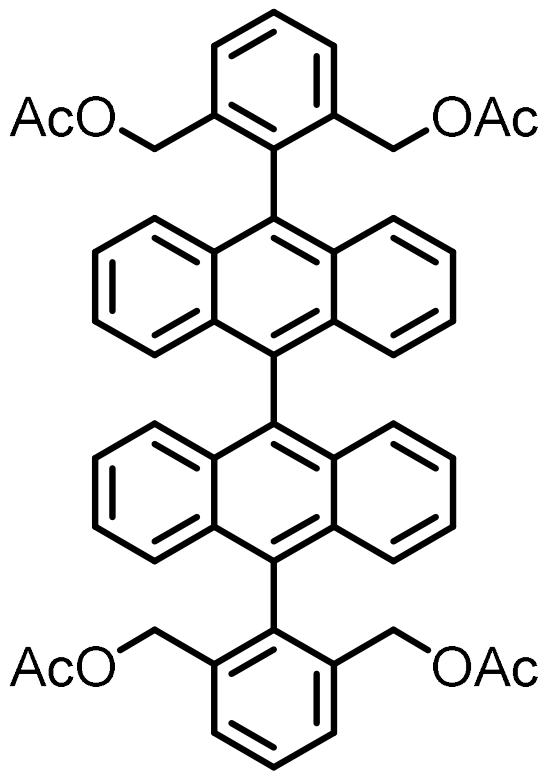

**5**

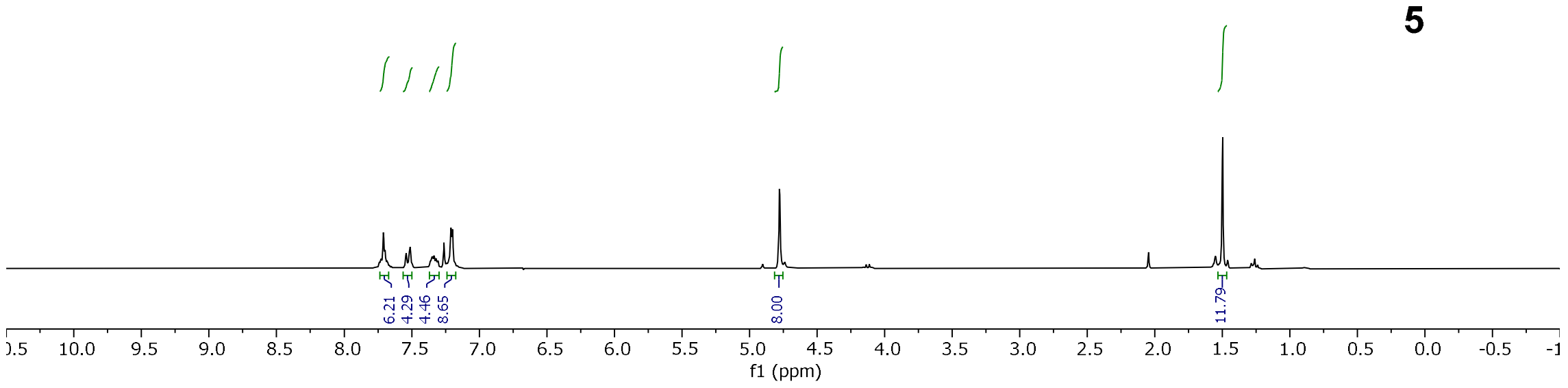

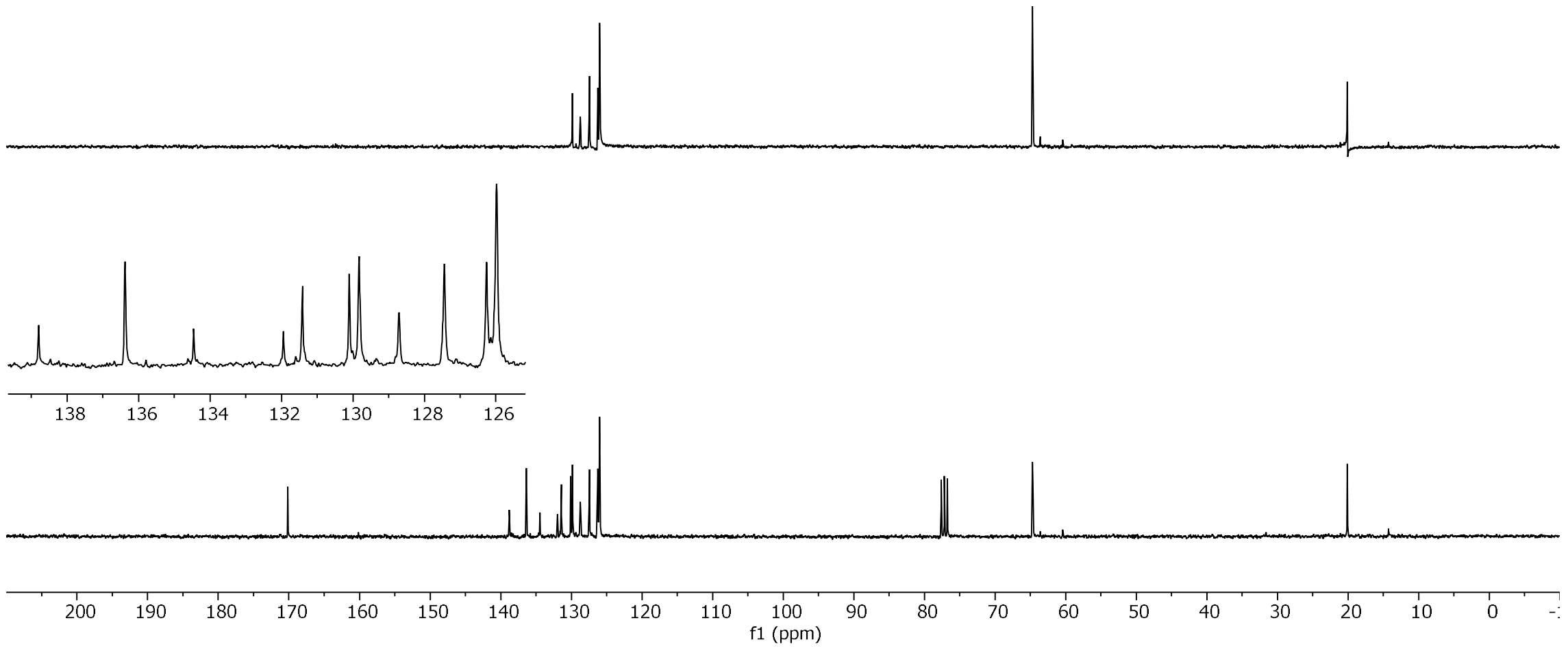

**Fig. S9.** $^{1}$H and $^{13}$C NMR spectra of **5**.

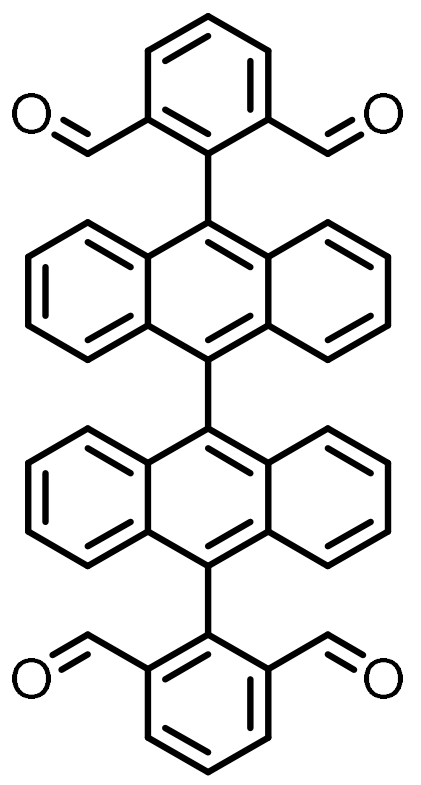

**6**

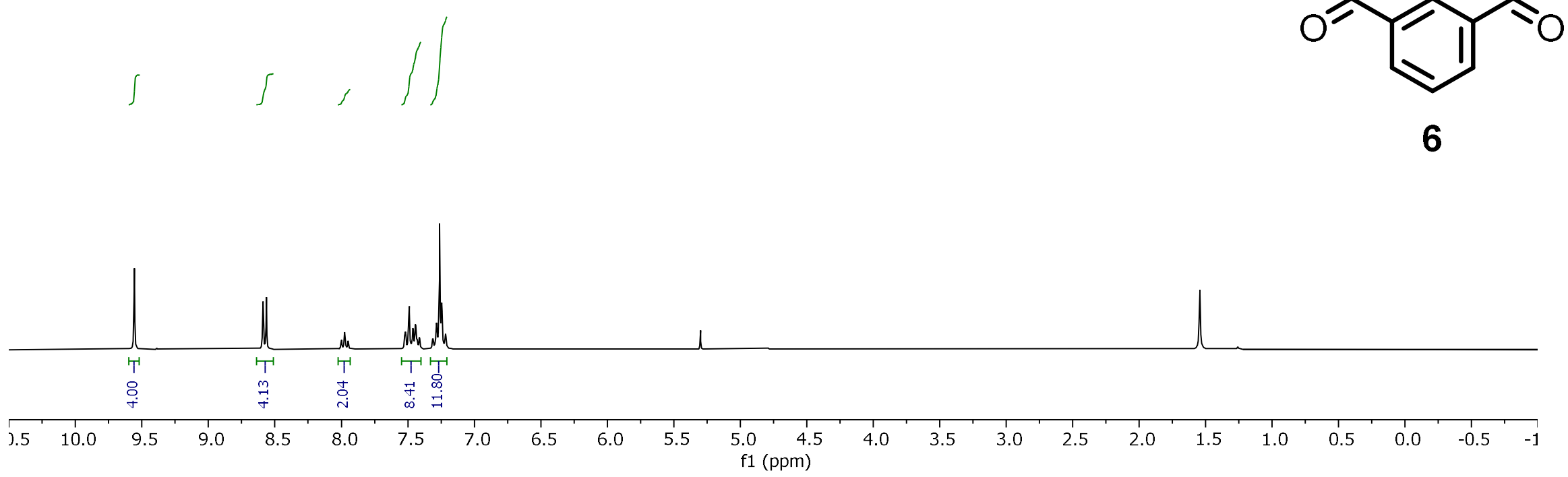

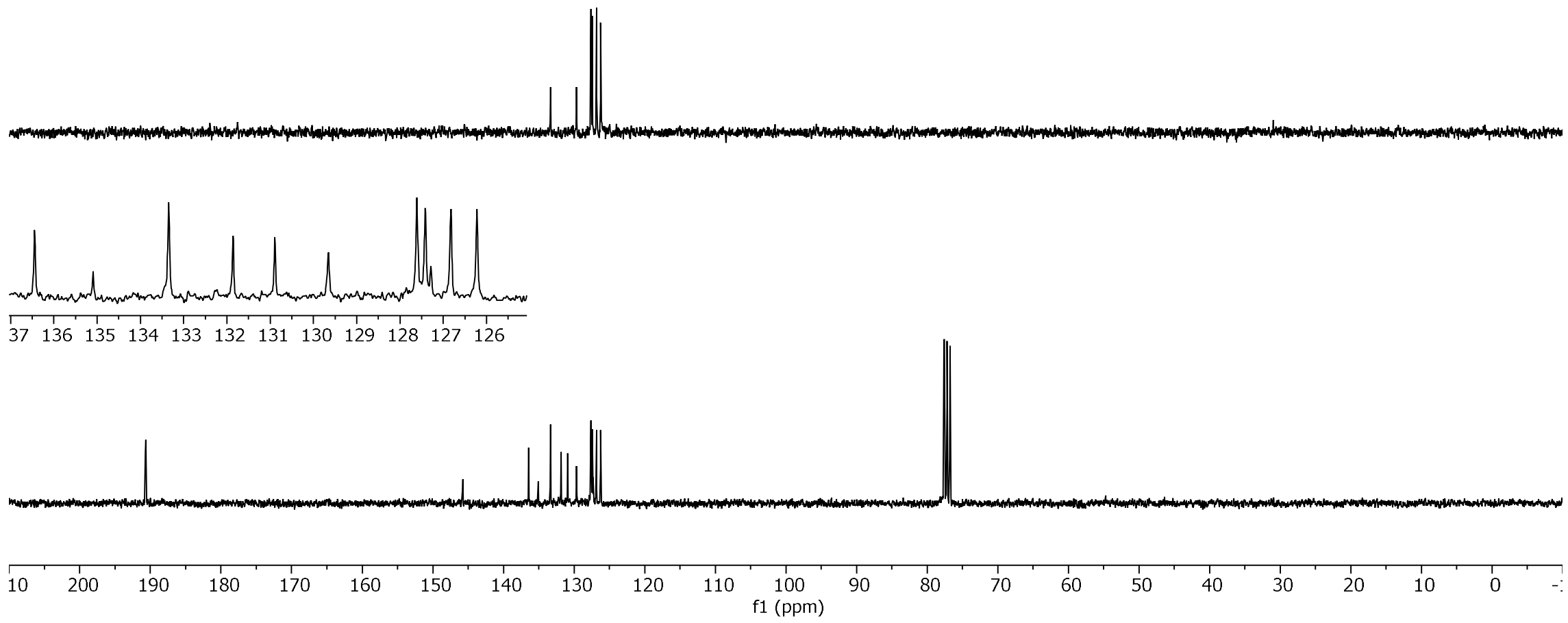

**Fig. S10.** $^{1}H$ and $^{13}C$ NMR spectra of **6**.

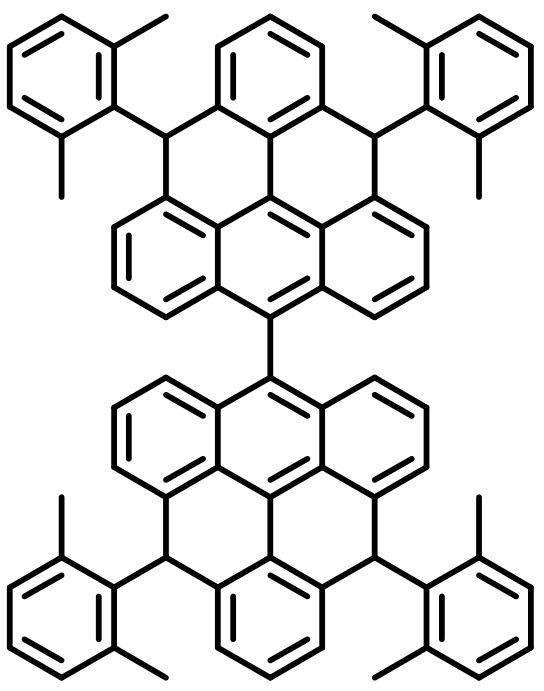

**1**

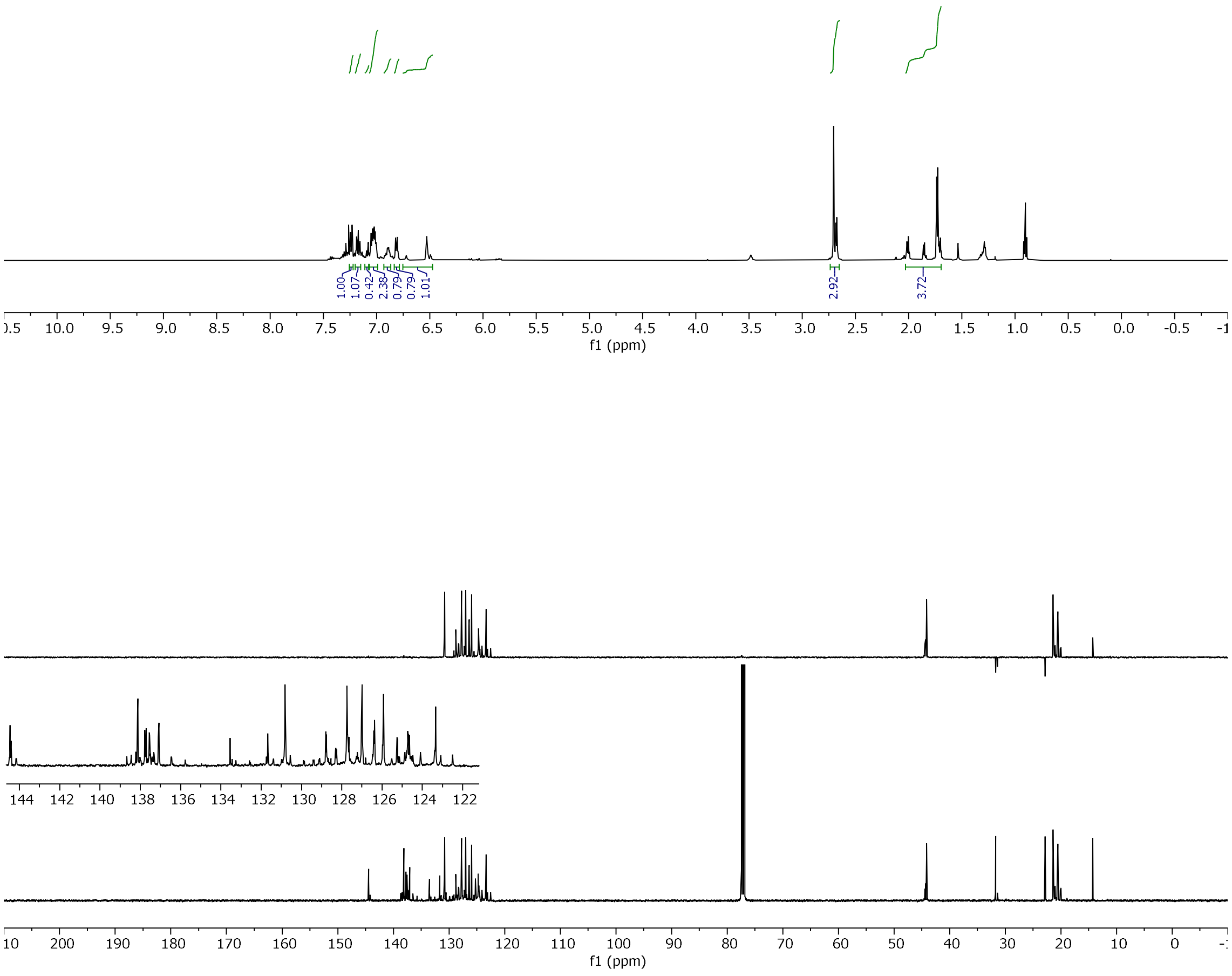

**Fig. S11.** $^1$H and $^{13}$C NMR spectra of **1**.

## 1.2. Sample preparation and scanning probe experiments.

Scanning probe measurements were performed in a home-built STM/AFM setup operating at base pressures below $1 \times 10^{-10}$ mbar and a base temperature of 5 K. Bias voltages were applied to the sample with respect to the tip. All STM and AFM images were acquired with carbon monoxide functionalized tips. AFM measurements were performed in non-contact mode with a qPlus[2] sensor. The sensor was operated in frequency modulation mode[3] with a constant oscillation amplitude of 0.5 Å. STM images were acquired in constant-current mode, and AFM images were acquired in constant-height mode with $V = 0$ V. Au(111) and Cu(111) surfaces were cleaned by iterative cycles of sputtering with $Ne^+$ ions and annealing up to 800 K. NaCl was thermally evaporated on Au(111) and Cu(111) surfaces held at 323 K and 283 K, respectively. This protocol results in the growth of predominantly bilayer (100)- terminated islands, with a minority of third-layer islands. Submonolayer coverage of **1** on the surfaces was obtained by flashing an oxidized silicon wafer containing the precursor molecules in front of the cold sample in the microscope. Carbon monoxide molecules for tip functionalization were dosed from the gas phase on the cold sample.

## 1.3. Tight-binding and mean-field Hubbard calculations.

Nearest-neighbor tight-binding and mean-field Hubbard calculations were performed by numerically solving the following Hamiltonian

$$\hat{H} = -t \sum_{\langle i,j \rangle,\sigma} c_{i,\sigma}^{\dagger} c_{j,\sigma} + U \sum_{i,\sigma} \langle n_{i,\sigma} \rangle n_{i,\bar{\sigma}} - U \sum_{i} \langle n_{i,\uparrow} \rangle \langle n_{i,\downarrow} \rangle. \quad (1)$$

Here, $c_{i,\sigma}^{\dagger}$ and $c_{j,\sigma}$ denote the spin selective ($\sigma \in \{\uparrow,\downarrow\}$ with $\bar{\sigma} \in \{\downarrow,\uparrow\}$) creation and annihilation operator at nearest-neighbor sites $i$ and $j$, $t$ = 2.7 eV is the nearest-neighbor hopping parameter, $U$ = 4 eV is the on-site Coulomb repulsion, $n_{i,\sigma}$ and $\langle n_{i,\sigma} \rangle$ denote the number operator and mean occupation number at site $i$, respectively. Orbital electron densities, $\rho$, of the $n^{\text{th}}$-eigenstate with energy $E_n$ have been simulated from the corresponding state vector $a_{n,i,\sigma}$ by

$$\rho_{n,\sigma}(\vec{r}) = \left| \sum_{i} a_{n,i,\sigma} \phi_{2p_z}(\vec{r} - \vec{r}_i) \right|^2, \quad (2)$$

where $\phi_{2p_z}$ denotes the Slater $2p_z$ orbital for carbon.

## 1.4. Density functional theory calculations.

**Gas-phase calculations.** The **ECG** geometry was optimized at the B3LYP/def2-SVP level of theory[4, 5] in the high-spin (quintet) state. Magnetic properties (NICS(1)$_{zz}$ and bond currents) were probed at this geometry and spin state at the B3LYP/def2-TZVP level of theory. These calculations were performed using Gaussian16,[6] with SYSMOIC[7] being employed for bond currents.

**On-surface calculations.** Calculations were done using VASP at the PBE-D3BJ level of theory.[8, 9] A 10×10 supercell of the Cu(111) surface was prepared from a face-centered lattice with the bulk lattice constant (3.614 Å), and 12 Å were added in the *z*-direction. Due to the large size of the supercell, only three layers of Cu atoms were included, and *k*-space was sampled only at the gamma point. The adsorption energy was estimated according to

$$E_{\text{ads}} = E_{\textbf{ECG@}\text{Cu111}} - E_{\textbf{ECG}} - E_{\text{Cu111}} \quad (3)$$

The value of $E_{\textbf{ECG}}$ was estimated by relaxing the high-spin (quintet) geometry of **ECG** in the gas phase, using a sufficiently large unit cell, and the surface energy $E_{\text{Cu111}}$ was determined by relaxing the uppermost layer of Cu atoms in the prepared supercell. Finally, $E_{\textbf{ECG@}\text{Cu111}}$ was obtained by placing **ECG** onto the reconstructed surface in four orientations (with the long axis of **ECG** rotated by ~0°, ~30°, ~60°, and ~90° with respect to the *a* lattice vector) and performing spin-polarized relaxations on the quintet manifold. Relaxations (with forces converged to 0.01 eV/Å) found the two lowest-energy adsorption configurations corresponding to ~90° and ~70° rotations of the long axis of **ECG** with respect to the *a* lattice vector (Fig. S15a,b), with the respective adsorption energies being –9.51 eV (Fig. S15a) and –9.38 eV (Fig. S15b), corresponding to an interaction of 123–125 meV per carbon atom. Ring puckering of 0.55–0.60 Å was observed in both configurations, with the spin-polarized edges being closer to the surface than the center of the molecule, in agreement with the AFM data (Fig. S14).

## 1.5. Multiconfigurational quantum chemistry calculations.

The solutions of a bilinear (BL) Heisenberg Hamiltonian between particles *i* and *j* coupled by $J_{ij}$

$$\hat{H}_{\mathrm{BL}} = -\sum_{i,j} J_{ij}\, \mathbf{S}_i \cdot \mathbf{S}_j \tag{4}$$

for a system of four spin-1/2 particles are two singlets ($S$ = 0), three triplets ($S$ = 1), and one quintet (S = 2). The energies of these states were obtained from first principles by a difference-dedicated configuration interaction (DDCI) calculation based on a complete active space (CAS) calculation with four electrons and four orbitals in an active space on the gas-phase optimized **ECG** geometry. The minimal basis set (STO-3G; one $p_z$ orbital per carbon) was used. DDCI was chosen due to its excellent performance in modeling exchange couplings.[10]

The energy spectrum of magnetic excitations obtained using DDCI is shown in the first column of Fig. S17c. To determine exchange couplings, DDCI results were fit to the BL Hamiltonian (eq. 4), with results shown in Fig. S17a and the second column of Fig. S17c. As the (ferromagnetic) coupling between electrons at the same side of **ECG** is much stronger than the coupling between electrons on the opposite sides ($|J_1| >> |J_2| \approx |J_3|$; see Fig. S17a), the low-energy spectrum of **ECG** can be also be described with the bilinear-biquadratic (BLBQ) Hamiltonian[11] applied to a two spin-1 particle system

$$\hat{H}_{\mathrm{BLBQ}} = \sum_{i,j} -J_{ij}\, \mathbf{S}_i \cdot \mathbf{S}_j - B_{ij}\left(\mathbf{S}_i \cdot \mathbf{S}_j\right)^2 \tag{5}$$

Where $J_{ij}$ is the bilinear and $B_{ij}$ the biquadratic coupling between spin-1 particles $i$ and $j$. Results of the BLBQ Hamiltonian are shown in Fig. S17b and the third column of Fig. S17c.

# 2. Supporting data

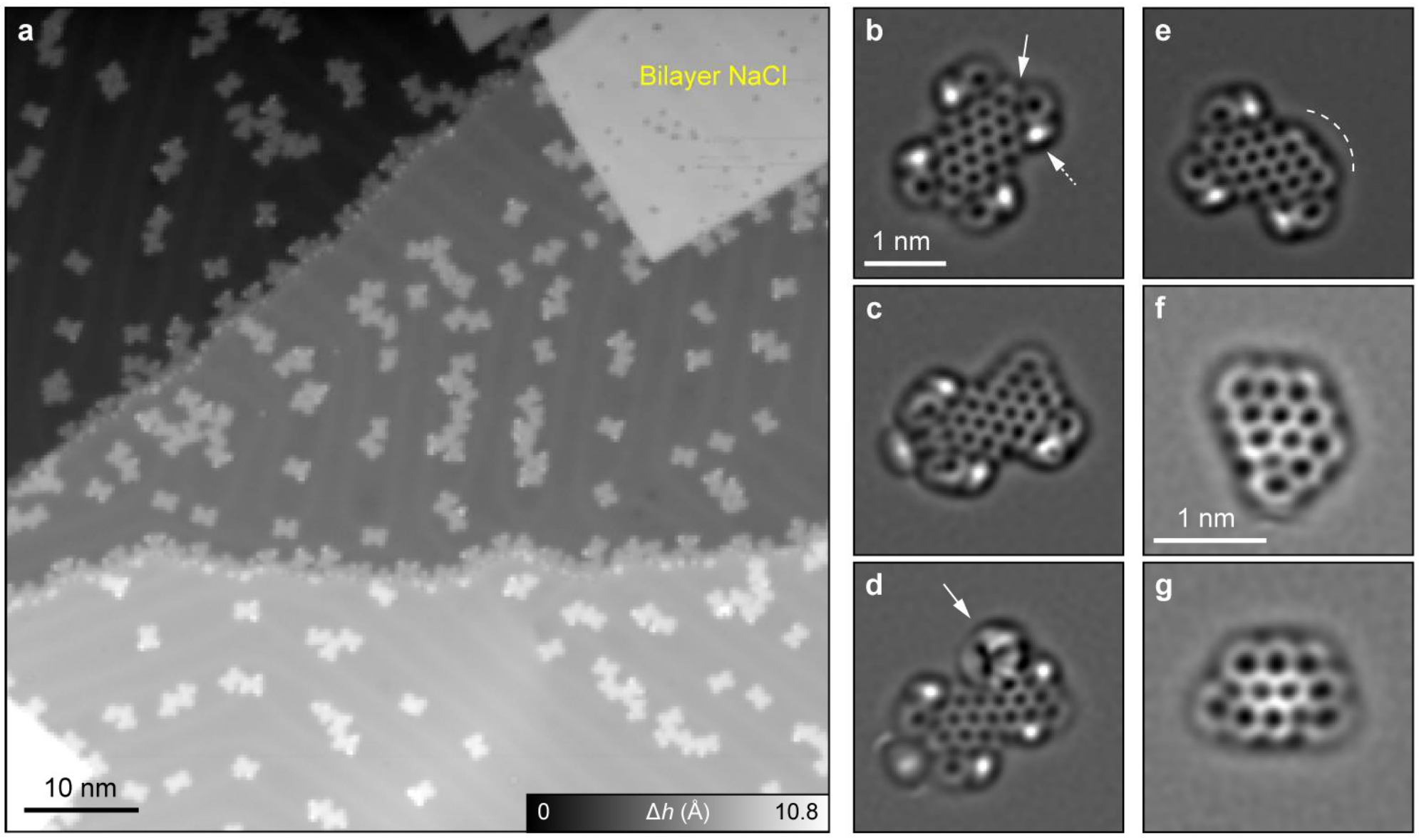

**Fig. S12.** On-surface reactions of **1** on Au(111). (a) Overview STM image after annealing **1** on Au(111) at 600 K (*V* = 0.2 V, *I* = 1 pA). (b–g) Laplace-filtered AFM images of several isolated molecules on the surface. We always observed loss of one or more methyl groups in **1** that led to the formation of pentagonal rings upon cyclodehydrogenation reactions, indicated by the solid arrow in (b). Where methyl groups are not lost and hexagonal rings are formed, we mostly observed incomplete dehydrogenation leading to methylene moieties, indicated by the dashed arrow in (b). The molecules in (b), (c) and (d) result from loss of four, three and two methyl groups, respectively. The molecule in (d) additionally contains an unreacted xylyl group indicated by the arrow. The molecule in (e) exhibits loss of three methyl and one xylyl groups (the dashed curve indicates the region from where the xylyl group is lost). The molecules in (f) and (g) result from fragmentation of the precursor in half, along with loss of one (f) and two (g) methyl groups. For AFM imaging, the tip was approached by 1.5 (b–f) and 1.7 Å (g) from the STM setpoint of *V* = 0.2 V, *I* = 1 pA on Au(111). The scale bar in (b) also applies to (c–e), and the scale bar in (f) also applies to (g).

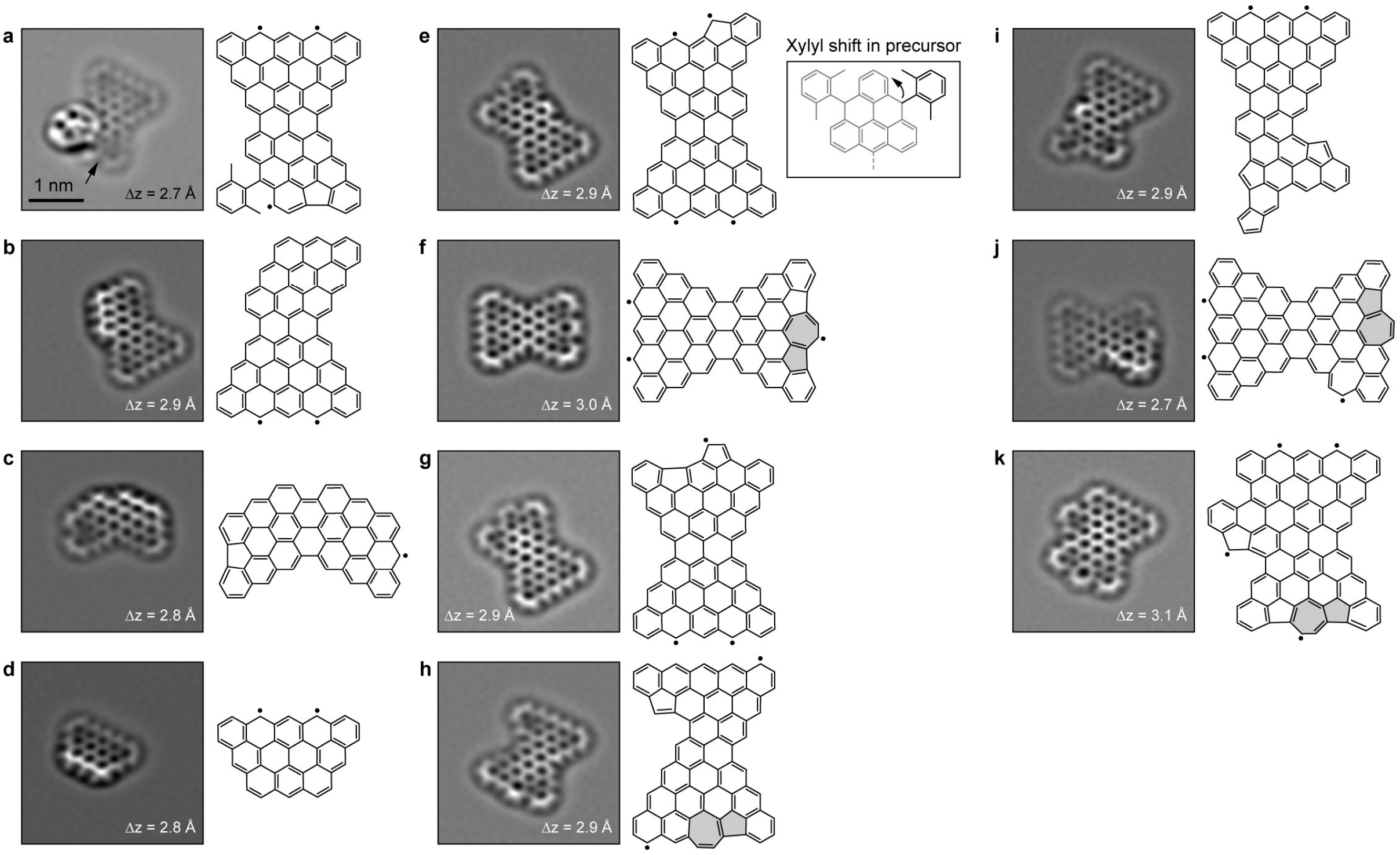

**Fig. S13.** (a–k) Laplace-filtered AFM images and corresponding chemical structures of selected molecules that do not correspond to **ECG**, obtained after annealing **1** on Cu(111) at 530 K. The molecules in (a–d) result from loss of methyl and/or xylyl groups (a–c), and precursor fragmentation (d), as also observed on Au(111) (Fig. S12). The molecule in (e) results from a xylyl shift in **1** and subsequent cyclization reactions. The molecules in (f–k) result from more complex chemical reactions. Some of them (g, j, k) contain more carbon atoms than in **ECG**, which could be explained by the reaction of molecular species with small fragments (such as methyl radicals). Other molecules contain azulene moieties (gray filled rings) that may result from ring rearrangement reactions. For AFM imaging, the tip was approached by a distance $\Delta z$ indicated in each panel from the STM setpoint of $V$ = 0.2 V, $I$ = 0.5 pA on Cu(111). The scale bar in (a) also applies to (b–k).

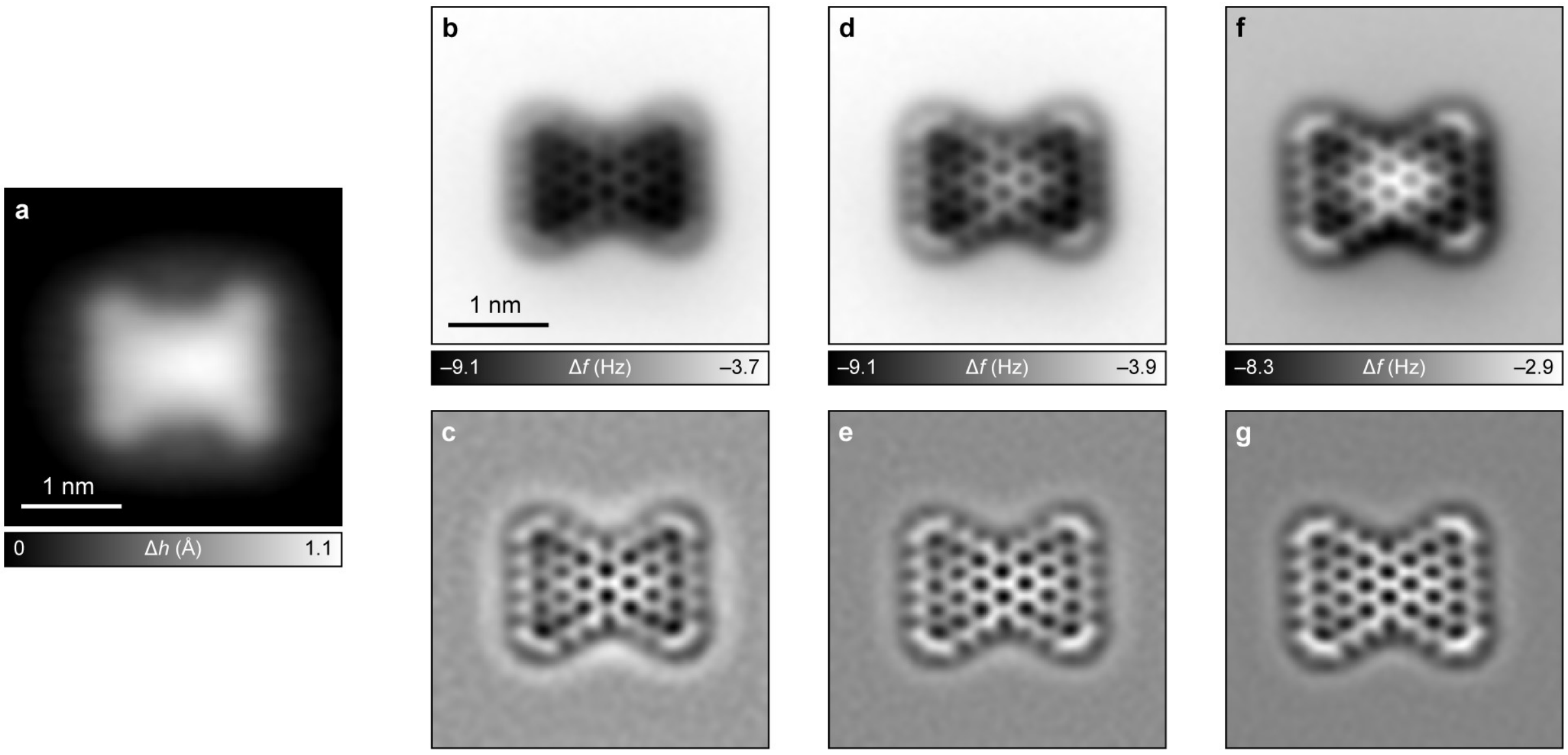

**Fig. S14.** Height-dependent AFM imaging of **ECG** on Cu(111). (a) STM image of **ECG** on Cu(111) ($V$ = 0.2 V, $I$ = 0.5 pA). (b–g) AFM images (b, d, f) and corresponding Laplace-filtered images (c, e, g) of **ECG** acquired at different tip heights. The tip was approached by 2.6 (b, c), 2.8 (d, e) and 3.0 Å (f, g) from the STM setpoint of *V* = 0.2 V, *I* = 0.5 pA on Cu(111). The scale bar in (b) also applies to (c–g).

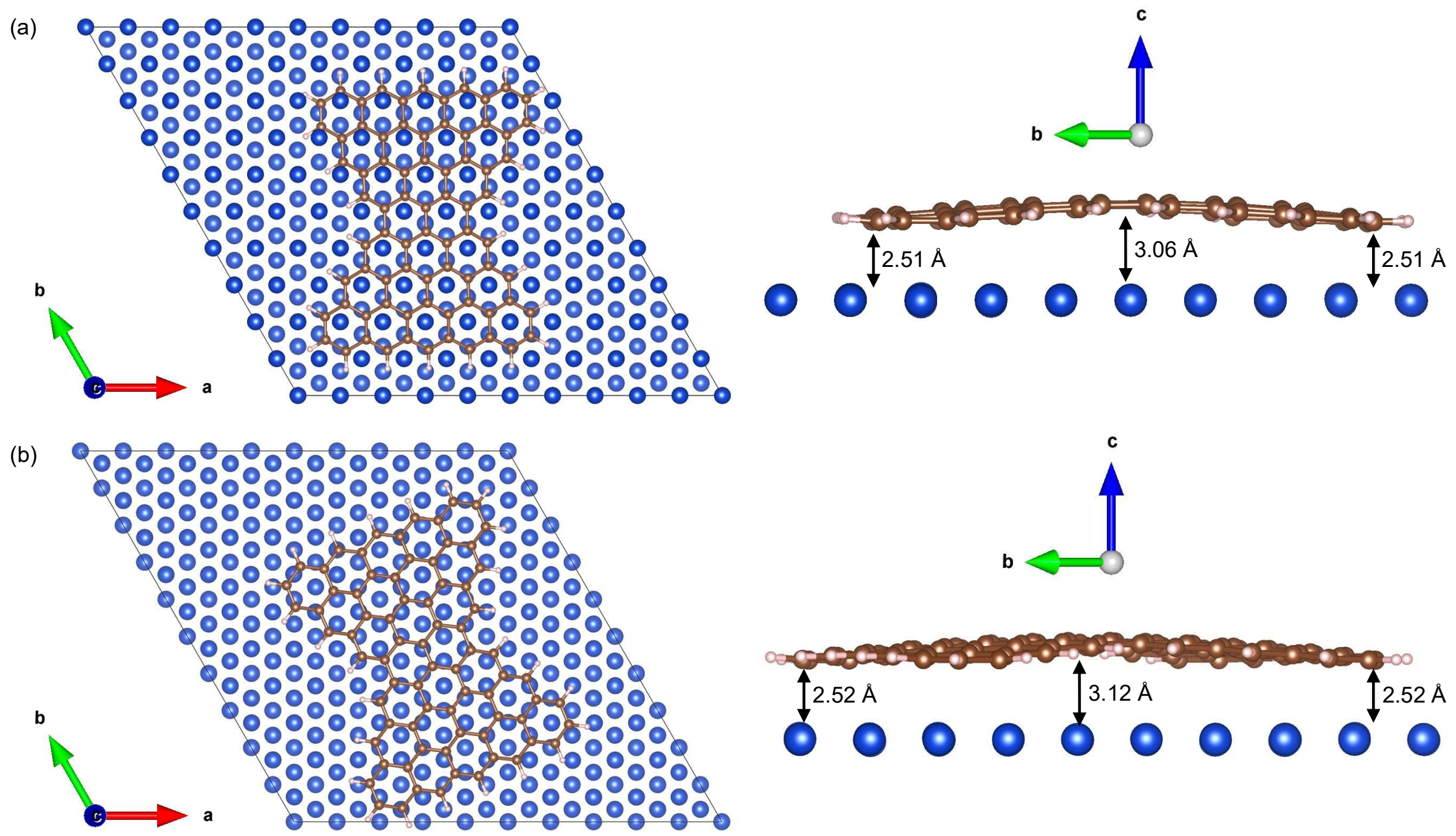

**Fig. S15.** Top (left) and side (right) views of the DFT-optimized geometries of **ECG** on Cu(111). The two lowest-energy configurations are shown, with the long axis of **ECG** rotated by (a) ~90° with respect to lattice vector *a* and $E_{ads}$ = –9.51 eV, and (b) ~70° with respect to *a* and $E_{ads}$ = –9.38 eV.

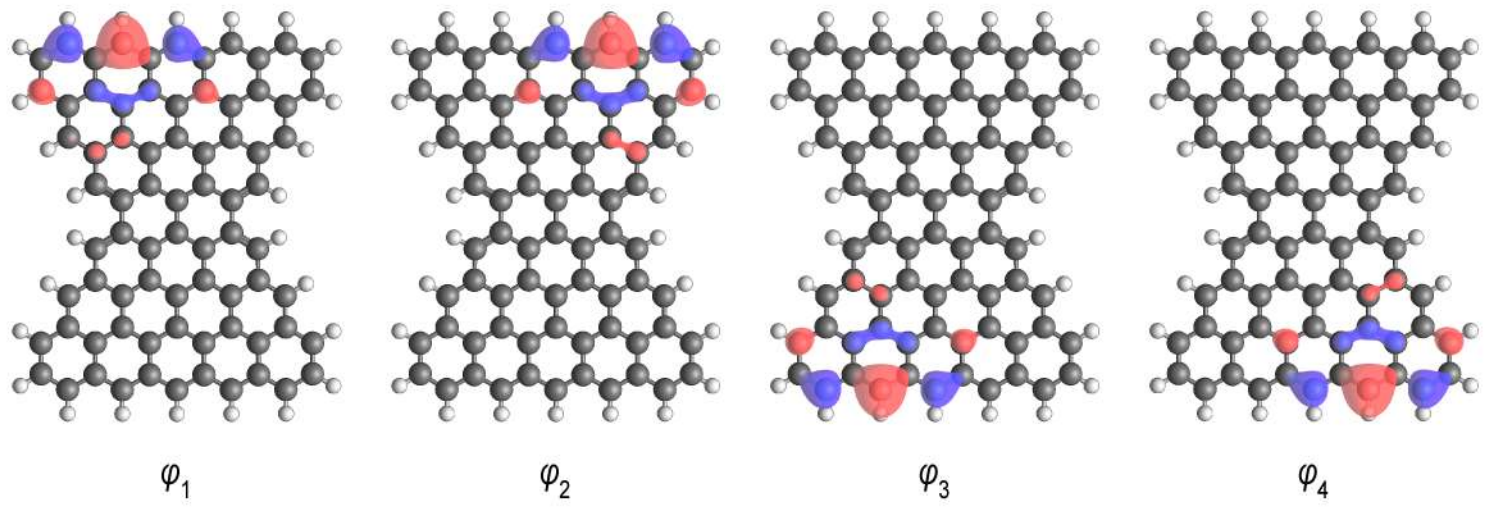

**Fig. S16.** CASSCF(4,4) localized orbitals of **ECG**.

**Supporting Note.** Here, we provide the composition of the CASSCF(4,4)-DDCI many-body ground and excited states of **ECG** in the localized orbital basis $|\varphi_1\varphi_2\varphi_3\varphi_4\rangle$ (Fig. S16).

$$S_0 = 0.58(|\uparrow\uparrow\downarrow\downarrow\rangle + |\downarrow\downarrow\uparrow\uparrow\rangle) - 0.29(|\uparrow\downarrow\uparrow\downarrow\rangle + |\uparrow\downarrow\downarrow\uparrow\rangle + |\downarrow\uparrow\uparrow\downarrow\rangle + |\downarrow\uparrow\downarrow\uparrow\rangle)$$

$$T_1 = 0.5(|\downarrow\uparrow\uparrow\uparrow\rangle + |\uparrow\downarrow\uparrow\uparrow\rangle - |\uparrow\uparrow\downarrow\uparrow\rangle - |\uparrow\uparrow\uparrow\downarrow\rangle)$$

$$Q_1 = |\uparrow\uparrow\uparrow\uparrow\rangle$$

$$T_2 = 0.5(-|\downarrow\uparrow\uparrow\uparrow\rangle + |\uparrow\downarrow\uparrow\uparrow\rangle - |\uparrow\uparrow\downarrow\uparrow\rangle + |\uparrow\uparrow\uparrow\downarrow\rangle)$$

$$T_3 = 0.5(-|\downarrow\uparrow\uparrow\uparrow\rangle + |\uparrow\downarrow\uparrow\uparrow\rangle + |\uparrow\uparrow\downarrow\uparrow\rangle - |\uparrow\uparrow\uparrow\downarrow\rangle)$$

$$S_1 = 0.50(|\uparrow\downarrow\downarrow\uparrow\rangle + |\downarrow\uparrow\uparrow\downarrow\rangle - |\uparrow\downarrow\uparrow\downarrow\rangle - |\downarrow\uparrow\downarrow\uparrow\rangle)$$

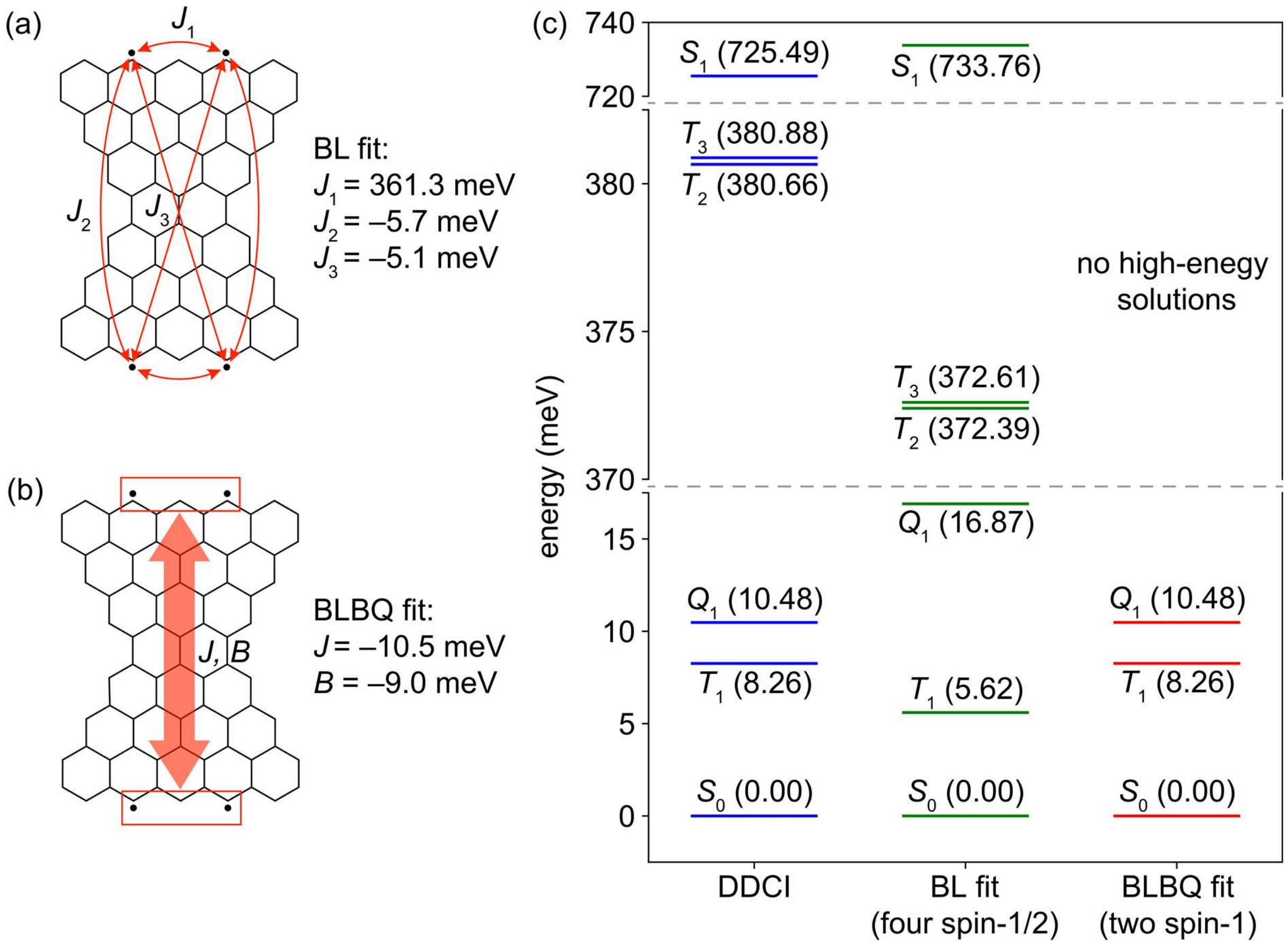

**Fig. S17.** Magnetic interactions in **ECG**. (a) The spin-1/2 bilinear (BL) and (b) the spin-1 bilinear-biquadratic (BLBQ) effective Hamiltonians, with exchange couplings and their fit to DDCI results shown. (c) Energy spectrum obtained from DDCI (blue, first column), and the fitted BL (green, second column) and BLBQ (orange, third column) Hamiltonians.